\documentclass{article}

\usepackage{arxiv}

\usepackage[utf8]{inputenc} 
\usepackage[T1]{fontenc}    
\usepackage{hyperref}       
\usepackage{url}            
\usepackage{booktabs}       
\usepackage{amsfonts}       
\usepackage{nicefrac}       
\usepackage{microtype}      
\usepackage{lipsum}
\usepackage{graphicx}
\usepackage{xcolor}
\usepackage{csquotes}
\usepackage{mathtools}
\usepackage{longtable}
\usepackage{pdflscape}
\usepackage{multirow}
\usepackage{tcolorbox}
\usepackage[disable]{todonotes}

\hypersetup{
    colorlinks=true,
    linkcolor=blue,
    filecolor=blue,      
    urlcolor=blue,
}

\title{The computational Patient has diabetes and a COVID}

\author{
  Pietro Barbiero\thanks{Use footnote for providing further
    information about author (webpage, alternative
    address)---\emph{not} for acknowledging funding agencies.} \\
  Department of Computer Science and Technology, University of Cambridge, Cambridge, UK \\
  \texttt{pietro.barbiero@tutanota.com} \\
   \And
 Pietro~Li\'o \\
  Department of Computer Science and Technology, University of Cambridge, Cambridge, UK \\
  \texttt{pl219@cam.ac.uk} \\
}

\begin{document}
\maketitle\todo{TITLE: "Diabetes and COVID-19: a computational patient simulation study" or "The Computational Patient: Simulating the interaction of diabetes and COVID-19"}

\begin{abstract}

Medicine is moving from a curative discipline to a preventative discipline relying on personalised and precise treatment plans. The complex and multi level pathophysiological patterns of most diseases require a systemic medicine approach and are challenging current medical therapies. On the other hand, computational medicine is a vibrant interdisciplinary field that could help move from an organ-centered approach to a process-oriented approach. The ideal computational patient would require an international interdisciplinary effort, of larger scientific and technological interdisciplinarity than the Human Genome Project. When deployed, such a patient would have a profound impact on how healthcare is delivered to patients. Here we present a computational patient model that integrates, refines and extends recent mechanistic or phenomenological models of cardiovascular, RAS and diabetic processes. Our aim is twofold: analyse the modularity and composability of the model-building blocks of the computational patient and to study the dynamical properties of well-being and disease states in a broader functional context. We present results from a number of experiments among which we characterise the dynamic impact of COVID-19 and type-2 diabetes (T2D) on cardiovascular and inflammation conditions.  We tested these experiments under different exercise, meal and drug regimens. We report results showing the striking importance of transient dynamical responses to acute state conditions and we provide guidelines for system design principles for the inter-relationship between modules and components in systemic medicine. Finally this initial computational Patient can be used as a toolbox for further modifications and extensions.

\end{abstract}

\keywords{computational Patient \and computational Medicine \and Systems medicine \and COVID \and T2D Diabete \and Cardiovascular model \and Blood Pressure model}

\section{Introduction}

Computational medicine is increasingly effective to understand and predict complex physiological and pathological conditions in scenarios of single organ disease to comorbidities. Both mechanistic and phenomenological models are important aspects of computational medicine. When we formulate hypotheses on the mechanisms (usually involving molecules) underlying the behaviour of the various endpoints of a process, we could build a mechanistic model; when we formulate hypotheses based on the empirical observations of a phenomenon, we could build a phenomenological model. Most models are actually a combination of the two and there are certainly overlaps between phenomenological modeling, statistical and machine learning.
Mechanistic and phenomenological modeling aim at reproducing the main features of a real system with the minimum number of parameters and still providing explainability, interpretability and often causality. The objective is to gain a better understanding of how each of the different components of a biomedical system contribute to the overall process, its emerging properties and the causality relation of the occurred events.
A mechanistic and phenomenological model could be formulated using ordinary or partial differential equations \cite{Ascolani2015}, stochastic processes \cite{Giampieri2011}, logic \cite{Despeyroux2019} or in terms of a tailor-made syntax which could facilitate formal analysis and verification \cite{Bartocci2016, Bartocci2012}. 
The dedicated modeler may introduce a series of models of a process at different scales, from the molecular level to the whole body level, or describing processes occurring in different organs under the same disease conditions. Although there is growing awareness of long range communications in the body - for instance the communicome \cite{Ray2007} or the gut-brain axis \cite{Sudo2004}, the integration of various models in order to capture the behavior at systems medicine level has not been pursued so much. 
Examples of such multi level communication are given by the extensive network of comorbidities. Comorbidity is the term used to address diseases, often chronic ones, co-occurring in the same individual. 
An important challenge is the homogenisation of models across multiple spatial and time scales, which requires cell-level models to be systematically scaled up to the tissue/organ level, and related asymptotic techniques for the analysis of multiple timescale problems, such as those arising in processes communications. 
The cardiovascular system is usually described using a cardio-centric view. As an example, the heart is considered as the only pump in the system. Other pumps are actually the skeletal muscle which returns blood from the periphery to the central circulation. Another pump is embedded in the elastic arteries that use elastic properties to propel the blood forward.
This system is subtly coupled with the cardiovascular-associated nervous system and the blood pressure control which include the regulated inputs from many other organs, most notably lungs, kidney and pancreas \cite{Achanta2020}.
Therefore, the concept of cardiovascular disease could be reformulated as a more complexly connected system and disease landscape, perhaps inclusive of comorbidities, which could allow a better patient stratification and prognosis and consequently better drug discovery. 

In particular, infectious diseases are good examples of the need of inter-organ and inter-process modeling approaches as a pathogen fitness may require colonising different body environments. A current example is given by the COVID-19 pandemia.
Diabetes is a frequent comorbidity; the Coronado study has shown that $29\%$ of the people with T2D infected with COVID-19 were intubated and $10.6\%$ die in one week \cite{Cariou2020}.
The mortality statistics show that fighting the COVID-19 pandemia requires a focus on comorbidities. Many of the older patients who become severely ill have evidence of underlying illness such as cardiovascular disease, kidney disease, T2D or tumours \cite{Wang2020}. \todo{What about obesity?}
They make the largest percentage of patient who cannot breathe on their own because of severe pneumonia and acute respiratory distress syndrome and require intubation: 
about a quarter of intubated coronavirus patients die within the first few weeks of treatment \cite{Richardson2020}.

\section{Objectives}

\begin{tcolorbox}[colback=blue!5!white,colframe=blue!75!black,title=Objective 1]
  Introduce modular and composable paradigms for the design of computational patients.
\end{tcolorbox}

In Sec. \ref{sec:tool} we propose a modular approach for the design of personalised computational physiology systems. The complexity underlying multifactorial diseases requires the introduction of multi-scale, extensible and adaptable models where modular principles are used to break organism complexity and composable criteria to select, link and combine different components in a hierarchical fashion.

\begin{tcolorbox}[colback=blue!5!white,colframe=blue!75!black,title=Objective 2]
  Show how our approach may help in disclosing cascade effects of comorbidities.
\end{tcolorbox}

In Sec. \ref{sec:exp} we illustrate a concrete example where personalised comorbid conditions' dynamics can be modeled and analysed using our framework.
We focus on developing an integrated computational system modeling ripple effects of comorbidities on blood pressure regulation. To this end, in Sec. \ref{sec:back} we revised the physiological background required to understand the main underlying biological processes involved in this mechanism. 
Building upon previous studies, we devise a customisable computational patient in the form of a computational tool composed of extended versions of three publicly available mathematical models describing the circulatory system \cite{neal2007subject}, type-2 diabetes \cite{Topp2000}, and renin-angiotensin system (RAS) \cite{pilvankar2019glucose}, one of the main pathways regulating inflammatory response and blood pressure. Respiratory failure is a key feature of severe COVID-19 and a critical driver of mortality; 10.6\% of all diabetic patients hospitalised die within one week. Hence, in Sec. \ref{sec:methods} we propose a set of equations modeling the impact of type-2 diabetes on blood vessels' stiffness and the influence of additional external factors which can be personalised according to patient's characteristics and lifestyle habits. We introduce a variety of such elements describing the repercussions on blood pressures caused by ageing, type-2 diabetes, viral infections like COVID-19, ACE inhibitor treatments, meals, and physical exercise.

\section{Computational tool} \label{sec:tool}

\subsection{From integration to modularity and composability}\todo{condense section?}
In recent decades, the interest and the scientific effort in developing integrated quantitative and descriptive computational systems modeling physiological dynamics has rapidly grown. By 1997 the Physiome Project \cite{hunter2002iups} and the EuroPhysiome Initiative \cite{jan2007europhysiome} have actively devised and organised rich collections of mathematical models describing the functional behavior of components of living organisms, such as organs, cell systems, biochemical reactions, or endocrine systems. Such a modular approach has been primarily used to reduce complexity by deconvolving\todo{dividing?} the human physiome into elementary subunits. Indeed, each computational module can be seen as a standalone biological entity describing one of the structures, processes, or pathways of the whole organism. Yet, modeling physiological interactions, multi-scale signalling, and comorbidities requires the combination of multiple components to build more sophisticated computational systems. Several approaches have been proposed where different mathematical models have been integrated into a single system in order to describe synergistic effects and emerging phenomena \cite{neal2007subject}. Despite being widely used and accepted, such system design paradigm often requires an overwhelming amount of work in merging multiple systems together, and in tuning and validating the integrated model. Besides, technological advances in computer science in the last twenty years have dramatically changed coding languages and paradigms. Hence, different research groups have developed their computational systems on many different coding platforms, frameworks, and libraries, including general-purpose languages like MatLab, Java, Python, C, but also special-purpose ones like JSim. The variety of implementation platforms combined with the mathematical effort required to merge many different systems is in conflict with the urgent need of user-friendly, extensible, and adaptable system design paradigms where components can be selected and assembled in various combinations to satisfy specific requirements. Personalised medicine requires the introduction of novel system design paradigms where modules break organism complexity and composable criteria are used to select and combine different components. Instead of merging, tuning, and validating the whole integrated system, each module could be tuned and validated independently. Composable criteria may allow researchers to primarily focus on multi-scale signalling between modules. Tuning and validation may apply just on inter-module signals which will make the overall system independent on module-specific implementation characteristics.

\begin{figure}[!ht]
    \centering
    \includegraphics[scale=0.3]{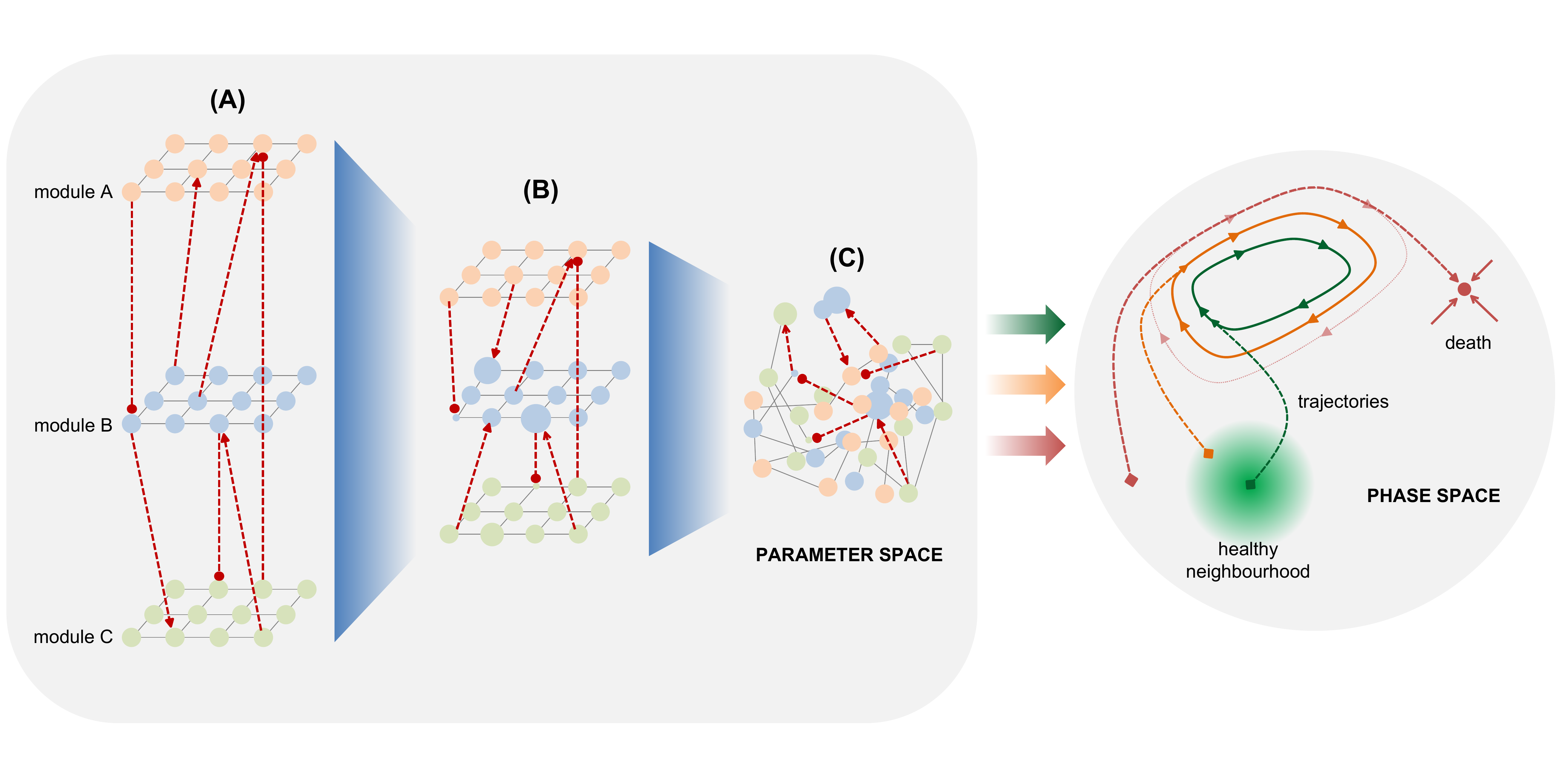}
    \caption{In modular systems several modules can be used independently to model physiological processes, disregarding their mutual relationships (\textbf{A}).
    The selection and combination of different components in a hierarchical fashion by means of composable criteria allows a better exploration of the parameter space (\textbf{B}). The actual interpenetration of multiple systems can be achieved by modeling the dynamics of their mutual relationships providing further information on the underlying phenomena (\textbf{C}).
    Such deeper exploration of the parameter space enhances the evaluation of initial conditions and trajectories in the phase space (\textbf{right}).}
    \small\flushleft Figure design inspired by \cite{gunawardena2010models}.
    \label{fig:model}
\end{figure}

\subsection{Module design and personalisation}
\begin{figure}[!ht]
    \centering
    \includegraphics[scale=0.45]{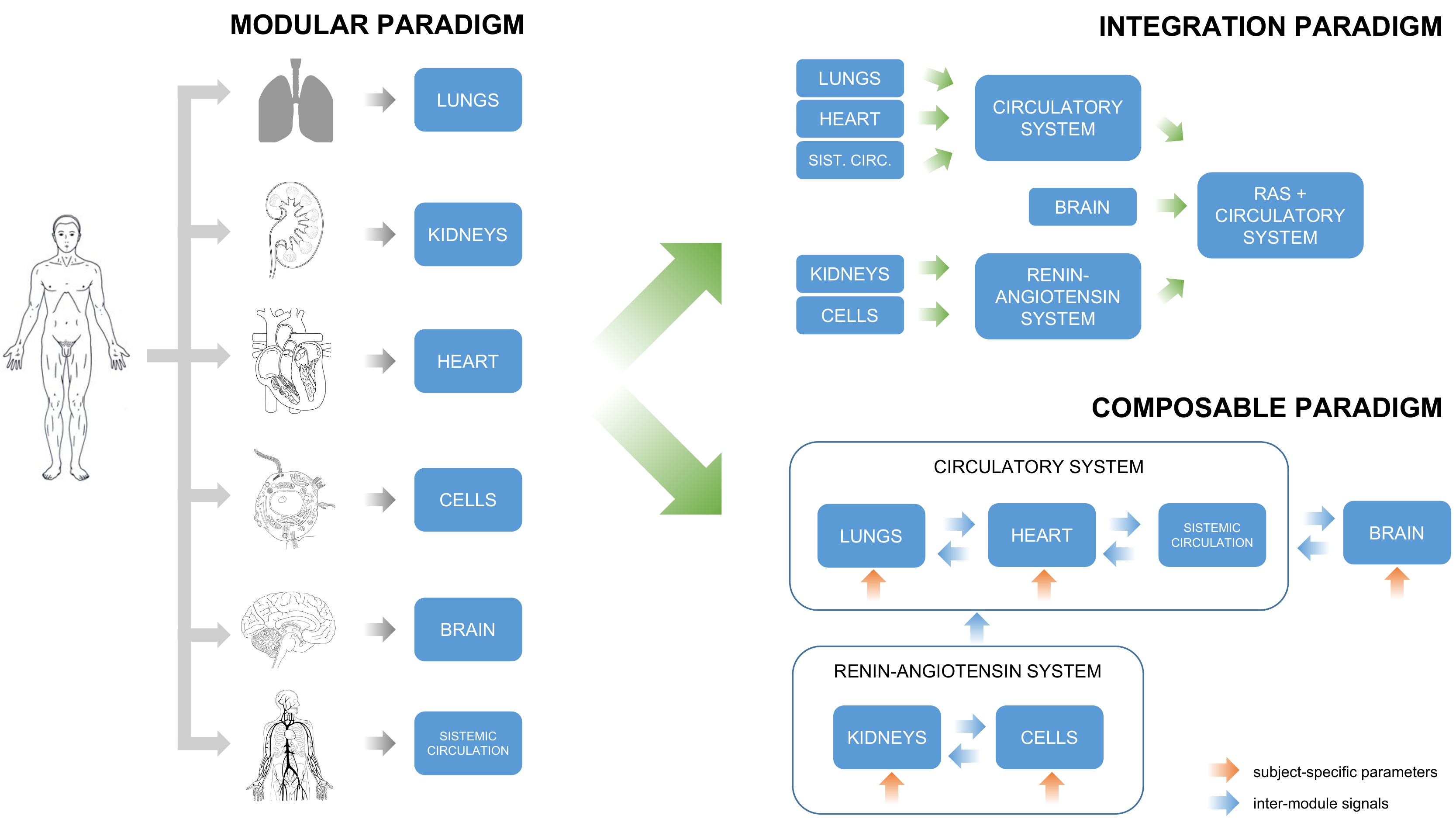}
    \caption{Modular paradigms are used to break organism complexity into simpler components which can be analysed and modeled independently (\textbf{left}).
    Integrating different modules requires an overwhelming amount of work in merging one after the other multiple systems together, and in tuning and validating the final model (\textbf{top right}).
    Composable criteria favor a dynamic and adaptable selection of different components allowing researchers to primarily focus on modeling the relationships between modules (\textbf{bottom right}).}
    \small\flushleft Illustrations adapted from \textit{The Sourcebook of medical illustration} \cite{cull1989sourcebook}.
    \label{fig:paradigm}
\end{figure}

In order to move towards this modern system design, each module can be seen as a black box processing signals coming from other modules and combining them with external subject-specific parameters in order to provide a set of responses (see Fig. \ref{fig:paradigm}). Subject-specific parameters may be derived from on-line clinically relevant measures, such as heart pressure or insulin levels, or from the Electronic Health Record \cite{kalra2006electronic,menachemi2011benefits}, such as morbidities, treatments, or clinical examinations. Such elements can be used to personalise the module taking into account unique subject characteristics. Incoming signals from other components may impact some of the variables and parameters of the module, but cannot change its architecture. Finally, the outputs provided by each module can be simultaneously used as inputs for other components or tracked as clinically relevant latent variables.

\subsection{Usage guidelines}\todo{reduce section?}
The computational system has been designed in order to allow for three levels of user interaction.
computational scientists and coders may take advantage of publicly available code by improving or forking the GitHub repository \cite{repo}. The repository structure has a modular design so that new packages can be included independently. Each new package should correspond to a new mathematical model. Multiple packages can be combined together in order to generate more complex computational systems. Medical practitioners and biologists with some Python experience may just download the repository, reproduce the simulations on their computers, or modify some parameters. In order to make the computational tool available for clinicians and practitioners without coding skills, the whole computational system has been incorporated into a website with a graphical user interface. Users may profit from this user-friendly interaction as the system can be customised in many different ways creating multiple scenarios by modifying several parameters, including patient-specific characteristics and constants related to models' interactions.

\subsection{Numerical methods}

All the necessary code for the experiments has been implemented in Python 3, relying upon open-source libraries.
The mathematical equations described in Sec. \ref{sec:methods} form a set of ODE systems and algebraic equations that
have been sequentially solved using the \texttt{LSODA} integration method  \cite{hindmarsh1983odepack,petzold1983automatic} provided by the function \texttt{solve\_ivp} included into the \texttt{scipy} Python package \cite{perez2009python}. 
All the experiments have been run on the same machine: Intel\textsuperscript{\textregistered} Core\texttrademark\ i7-8750H 6-Core Processor at 2.20 GHz equipped with 8 GiB RAM.

\section{Physiological background}
\label{sec:back}

The objective of our model is showing how the combined effects of comorbidities may lead to severe cardiovascular and pulmonary conditions. To this aim, we include in our model some of the main factors, pathways, and morbidities affecting blood pressure with a focus on pulmonary vessels, i.e. oxygenation, arterial stiffness, diabetes, RAS, and COVID-19. In this section we revise the physiological background of the elements involved in our computational system.

\subsection{The link between hypertension, oxygenation and blood pressure variability}

Exposure to chronic hypoxia causes pulmonary hypertension and pulmonary
vascular remodelling \cite{Howell2003}. 
COVID results in decreased oxygen that can result in impaired functioning of the heart and brain and cause difficulty with breathing (a PaO2 reading below 80 mm Hg or a pulse ox (SpO2) below 95 percent is considered low).
When the left side of the heart cannot pump blood out to the body normally, blood backs up in the lungs and increases blood pressure there.
The COVID-19 virus can activate the blood clotting pathway. Studies have reported  that 30\% of COVID-19 patients showed signs of blood clots in their lungs which means that a blood clot that has traveled to the lung.
One of the recommendations is to give a low dose of heparin, which prevents clot formation or tissue plasminogen activator (tPA), which helps to dissolve blood clots \cite{Sardu2020,Tang2020}.
High blood pressure can damage the arteries by making them less elastic,
which decreases the flow of blood and oxygen and leads to heart disease.
The relationship between blood pressure and stroke recurrence is controversial.
Recent researches stress that both high mean value of blood pressure and blood pressure variability (particularly long term) are important.
Although some variation in blood pressure throughout the day is normal, higher  variation in blood pressure is associated with a higher risk of cardiovascular disease and all-cause mortality \cite{Kim2018,Tao2017}. In young people here is a an increased blood supply response to hypoxia which could vanish in elderly with high blood pressure. This compromised response may be caused by the high blood pressure-induced impairment in the function of the blood vessels \cite{Fernandes2018}.

\subsubsection{Arterial stiffness}

Arterial stiffness is a broad term used to describe loss of arterial compliance and changes in vessel wall properties. 
Both arterial stiffness and high blood pressure variability can be indicators of cardiovascular risk \cite{Wen2015,ORourke2005,Mitchell2010,Bangalore2020,Clark2019}.
Ageing increases arterial stiffness and that increased arterial stiffness gives rise to increased blood pressure variability \cite{Uejima2019}.
Although arterial stiffness can be assessed using a variety of techniques, carotid–femoral pulse wave velocity is the preferred measure.
It has been shown that increased arterial stiffness is an early risk marker for developing type-2 diabetes \cite{Muhammad2017}, and a causal association between T2D and increased arterial stiffness has been proved on a large cohort of patients \cite{Xu2016,Gan2019}: 1 standard deviation increase in T2D is associated with 6\% higher risk in increased arterial stiffness; see also \cite{Eales2016}.  Arterial stiffness is also related to Inflammageing which is a chronic low-grade inflammation that develops with advanced age. It is believed to accelerate the process of biological ageing and to worsen many age-related diseases \cite{FRANCESCHI2006,Rea2018}.  In particular inflammatory cytokines (which may be activated by angiotensinII) result in increased arterial stiffness; on the contrary reductions in inflammation (for example due to anti-inflammatory cytokines), exercise reduce arterial stiffness \cite{Park2012,Madden2009}.

\subsection{The renin-angiotensin system and SARS-CoV-2}
The renin-angiotensin system (RAS) is a hormone system regulating vasoconstriction and inflammatory response \cite{fountain2019physiology}. The key regulator of the RAS is the peptide hormone Angiotensin II (ANG-II) generated by the angiotensin-converting enzyme (ACE) which cleaves the decapeptide Angiotensin I (ANG-I), or proangiotensin. ANG-II exerts its biological functions through two G-protein-coupled receptors, the ANG-II receptor type 1 receptor (AT1R) and ANG-II receptor type-2 receptor (AT2R), and the heptapeptide Angiotensin (1-7) (ANG-(1-7)) which binds and activates the G-protein-coupled Mas receptor (MAS). ANG-(1-7) can be generated both by the angiotensin-converting enzime 2 (ACE2) from ANG-II, or by the neutral endopeptidase enzyme (NEP) from ANG-I. The three G-protein-coupled receptors (AT1R, AT2R, and MAS) are the main factors helping the body to carry out the role of ANG-II in regulating blood pressure over the course of the day \cite{kuba2010trilogy}\cite{gironacci2011angiotensin}.
On one side, AT1R stimulates vasoconstriction, hypertension, and inflammatory response. The effect of AT1R is counterbalanced by MAS, promoting vasodilation, hypotension, and vasoprotection. The role of AT2R is currently debated \cite{matavelli2015at2}. Under normal physiologic conditions, AT2R counteracts most effects of AT1R. However, recent developments have shown how its vasodilatory effects were not associated with significant reduction in blood pressure \cite{foulquier2012impact}. In the kidney, AT2R stimulation produced natriuresis, increased renal blood flow, and reduced tissue inflammation \cite{kaschina2008angiotensin,gelosa2009stimulation,matavelli2011angiotensin}.
External factors impacting the RAS include: glucose concentration, ACE inhibitor treatements, and viral infections binding to ACE2, such as SARS-CoV-2. Glucose concentration has a direct impact both on AT1R and ACE activity. A high glucose concentration may determine chronic hypertensive conditions. Therefore, hypertensive treatments usually include ACE inhibitor drugs which are used to compensate the overproduction of ANG-II and AT1R \cite{zaman2002drugs}. Viral infections such as COVID-19 may also have a negative impact on RAS, as the virus binds to ACE2 in order to gain entry into the host cell, impairing the activity of ACE2 in generating ANG-(1-7) by hydrolyzing ANG-II \cite{south2020controversies}.

\begin{figure}[!ht]
    \centering
    \includegraphics[scale=0.5]{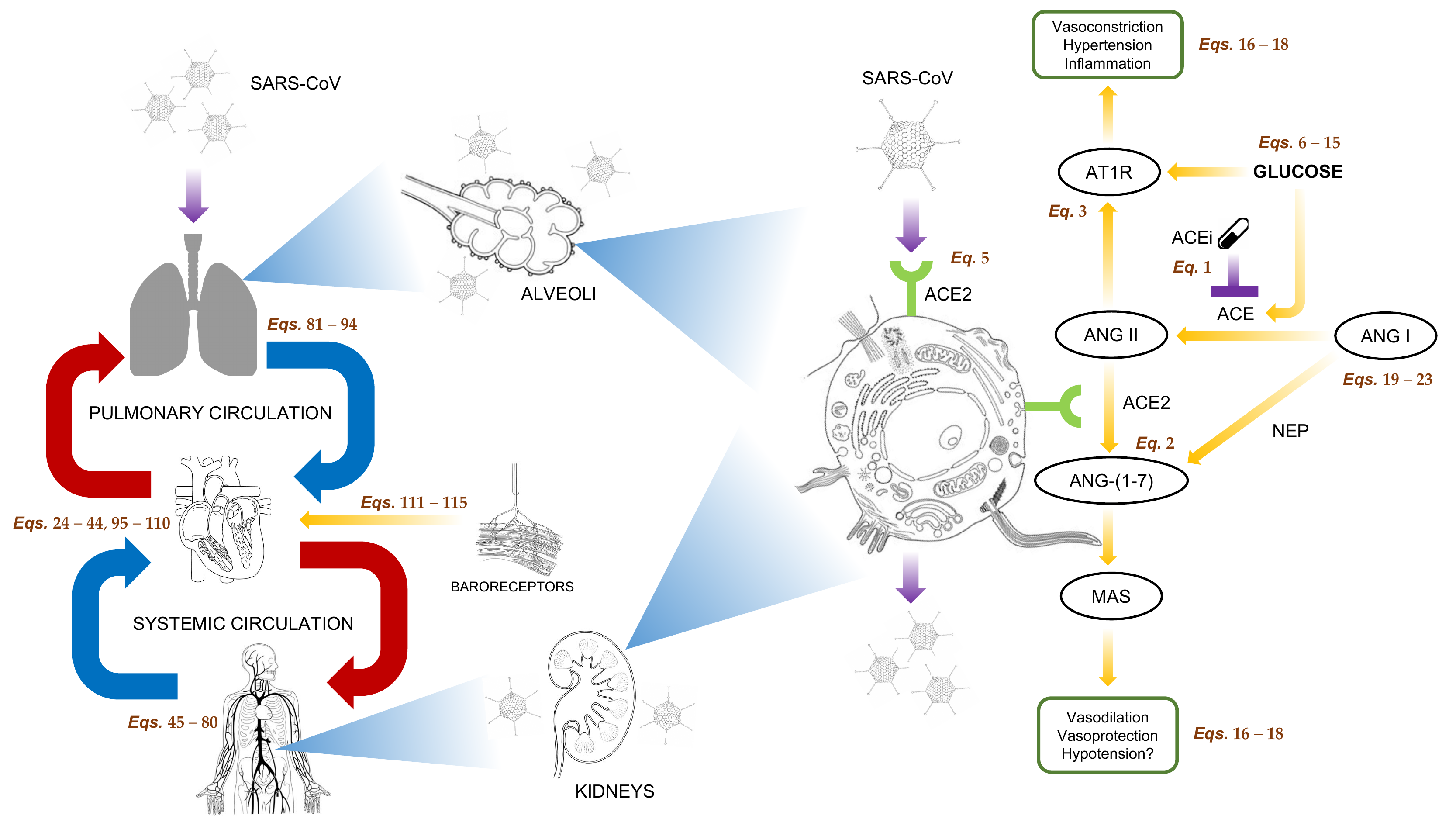}
    \caption{Schematic representation of the circulatory system composed of heart, pulmonary circulation, systemic circulation, and baroreceptors (\textbf{left}). External factors affecting the renin-angiotensin system (ACEi and SARS-CoV-2) are shown in {\color{violet} violet} (\textbf{right}).} \small\flushleft Illustrations adapted from \textit{The Sourcebook of medical illustration} \cite{cull1989sourcebook}.
    \label{fig:schema}
\end{figure}

\section{Mathematical model of diabetic computational patients} \label{sec:methods}

In this section we present a concrete example describing a set of mathematical models that can be used to model a computational patient.

We focus on modeling a diabetic computational patient by combining four modules: RAS \ref{sec:ras}, diabetic \ref{sec:diabetes}, circulatory \ref{sec:cardio}, and stiffness \ref{sec:stiff} models. Fig. \ref{fig:schema} shows a schematic representation of the computational system. The computational patient can be customised in two different ways. First, the system has been designed in order to be personalised using patient-specific values for some parameters such as age, glucose levels, arterial blood pressure, presence of comorbidities or treatments (see Table \ref{tab:custom}). Should the physiological analysis require the inclusion of additional conditions, new modules can be included and composed according to patient's needs.

\begin{table}[!ht]
\renewcommand{\arraystretch}{1.5}
\centering
\caption{computational patients' customisable parameters.}
\label{tab:custom}
\resizebox{\textwidth}{!}{
\begin{tabular}{lllll}
\hline
\textbf{Class} & \textbf{Parameter} & \textbf{Description} & \textbf{Values} & \textbf{Units} \\ \hline
\multirow{4}{*}{Clinical record}
 & $A$ & Age & $20-70$ & years \\
 & $G$ & Blood glucose levels & $100-200$ & $ml/dl$ \\
 & $ABP$ & Arterial blood pressure (from clinical records, e.g. \cite{neal2007subject}) & $80-120$ & $mmHg$ \\ 
 & $[D]$ & Vitamin D concentration  & $20-40$ & $ng/mL$ \\
\hline
\multirow{2}{*}{Comorbidities}
 & $infection$ & Presence/absence of SARS-CoV-2 infection & $\{\text{True}, \text{False}\}$ & - \\
 & $renal$ & Normal/impaired renal function & $\{\text{True}, \text{False}\}$ & - \\ 
\hline
\multirow{4}{*}{Treatments}
 & $drug$ & Presence/absence of ACEi treatments & $\{\text{True}, \text{False}\}$ & - \\
 & $d$ & ACEi dosage (Benazepril) & $0-5$ & $mg$ \\
 & $n_d$ & Number of daily drug administrations & $\{0,1\}$ & - \\
 & $[heparin]_0$ & Initial heparin dosage by intra venous injection & $5000-10000$ & $U/mL$ \\
\hline
\multirow{4}{*}{Lifestyle}
 & $t_w$ & Daily workout starting time & $6$ p.m. & - \\
 & $z_w$ & Daily workout intensity (burned calories) & $200$ & $kcal$ \\
 & $t_m$ & Daily meals' starting time & $\{8\text{ a.m. }, 12\text{ p.m. }, 8\text{ p.m.}\}$ & - \\
 & $g_m$ & Daily meals' glycemic load & $\{4, 42, 42\}$ & - \\
 & $s_m$ & Daily meals' carbohydrate serving & $\{50, 100, 100\}$ & $g$ \\
\hline
\end{tabular}
}
\end{table}


\subsection{Renin-angiotensin system and blood pressure regulation} \label{sec:ras}
The biochemical reaction network used to model the renin-angiotensin system is shown in Fig. \ref{fig:schema}. 
External factors include hypertension treatments and viral infections binding to ACE2, such as SARS-CoV-2. 
Hypertension drugs usually target ACE inhibiting ANG II production. ANG II promotes vasoconstriction, hypertension, inflammation, and fibrosis by activating AT1R. Therefore, reducing ANG II production with ACE inhibitors increases vasodilation and vasoprotection effects stimulated by AT2R and ANG-(1-7).
On the other hand, SARS-CoV-2 infections reduce ANG-(1-7) and ANG-(1-9) production rate, by binding to ACE2 in order to gain entry into the host cell. Hence, vasoprotection effects promoted by ANG-(1-7) decline, possibly leading to hypertension and inflammatory response.


\subsubsection{Pharmacokinetic model}
Pharmacokinetic (PK) models are used to describe drug absorption and excretion dynamics. Equation \ref{eq:pk} describes the analytical solution of a single-compartment pharmacokinetic model with first-order absorption and first-order elimination rates after oral administration \cite{versypt2017pharmacokinetic}. The equation has been used to model ACE inhibitors' dynamics and their effects on the RAS. A uniform dose size $d$ at constant time intervals $\tau$ has been assumed \cite{byers2009pharmacokinetic}:

\begin{equation} \label{eq:pk}
    [Drug]_n (t') = d\frac{k_a F}{(k_a - k_e) V} \Bigg(\frac{1 - \exp(-n k_e \tau)}{1 - k_e \tau} \exp{-k_e t'} - \frac{1 - \exp(-n k_a \tau)}{1 - k_a \tau} \exp{-k_a t'} \Bigg)
\end{equation}

where $[Drug]_n (t')$ is the drug concentration after the $n$-th dose, $t' = t (n-1) \tau$ is the time after the $n$-th dose, $k_a$ and $k_e$ are the absorption and elimination rates respectively, $F$ is the absorbed fraction of the drug, and $V$ the volume of distribution.

Pharmacokinetic parameters have been reported in table \ref{tab:pk}.


\subsubsection{Pharmacodynamic model}
Pharmacodynamic models are used to illustrate the effects of drug treatments on the body. 
The pharmacodynamic model used to describe local RAS dynamics has been derived from \cite{pilvankar2018mathematical, pilvankar2019glucose} (see Eqs. \ref{eq:drug}-\ref{eq:ANG2}).
The original model has been extended with four additional equations (Eqs. \ref{eq:ang17}-\ref{eq:ace2}). The variation of $[ANG17]$, $[AT1R]$ and $[AT2R]$ have been included as their dynamics can be useful in understanding how RAS regulates blood pressure \cite{lo2011using}. 
The concentration of ANG-(1-7) depends on the activity of two enzymes, NEP and ACE2, cleaving ANG-I and ANG-II respectively. $[AT1R]$ and $[AT2R]$ rather depend on $[ANGII]$ and on glucose concentration $G$.

\begin{eqnarray} \label{eq:ang17}
    \frac{d[ANG17]}{dt} &=& 
    \overbrace{k_{NEP} [ANG I]}^{\textrm{NEP-catalyzed conversion of ANG I}}
    + \overbrace{k_{ACE2} [ANG II]}^{\textrm{ACE2-catalyzed conversion of ANG II}}
    - \overbrace{\frac{\ln{2}}{h_{ANG17}} [ANG17]}^{\textrm{degradation}}
\end{eqnarray}

\begin{eqnarray}
    \frac{d[AT1R]}{dt} &=& 
    \overbrace{\Big( a_{AT1R} G + b_{AT1R} \Big) [ANGII]}^{\textrm{ANG-II bounds}}
    - \overbrace{\frac{\ln{2}}{h_{AT1R}} [AT1R]}^{\textrm{degradation}}
\end{eqnarray}

\begin{eqnarray} \label{eq:at2r}
    \frac{d[AT2R]}{dt} &=& 
    \overbrace{k_{AT2R} [ANGII]}^{\textrm{ANG-II bounds}}
    - \overbrace{\frac{\ln{2}}{h_{AT2R}} [AT2R]}^{\textrm{degradation}}
\end{eqnarray}

The dynamics of ACE2 activity ($k_{ACE2}$) has been introduced as an indicator of SARS-CoV-2 infectivity \cite{south2020controversies}:

\begin{eqnarray} \label{eq:ace2}
    \frac{d k_{ACE2}}{dt} = 
    \begin{dcases}
        s_V [ANGII] - e_{AI} k_{ACE2} &\qquad \text{during SARS-CoV-2 infection}\\
        0 &\qquad \text{otherwise}
    \end{dcases}
\end{eqnarray}

where $s_V$ represents the severity of the viral infection and $e_{AI}$ the efficiency of anti-inflammatory pathways. A higher concentration of $[ANGII]$ may also induce cells to produce more ACE2, thus increasing its activity \cite{south2020controversies} and enhancing viral entry. Hence ACE-inhibitor treatments may have a protective role as they reduce ACE activity lowering ANG-II levels (see Fig. \ref{fig:schema}).

Pharmacodynamic parameters and initial conditions have been reported in table \ref{tab:pd}.

\subsection{Adding comorbidities: type-2 Diabetes}
\label{sec:diabetes}

type-2 diabetes is a metabolic disease whose progression and severity is caused by increasing failure of insulin-production due to beta cell death.

There are complex multifactorial links between diabetes and cardiovascular disease \cite{Sharif2019,deOliveiraAlvim2013,Lu2020,Diaz2017}. The  main pathophysiology cornerstone is a state of chronic, low-level inflammation. This immune activation may facilitate both the insurgence and progression of insulin resistance in diabetic and pre-diabetic states and increases their cardiovascular risk. 
An extension of a model from Topp and collaborators (Eqs. \ref{eq:glucose}-\ref{eq:insulin-resistance}) combines insuline resistance, functional $\beta$-cell mass dynamics with glucose dynamics and insulin dynamics \cite{Topp2000}. The insulin and glucose dynamics are faster than the beta cell dynamics. Mild hyperglycaemia leads to increasing beta cell numbers, but above a threshold of $250mg/dL$ blood glucose, beta cell death is greater than cell division.
Additional terms (not shown) include and non-functional beta cells ($\beta _{f}$ and $\beta _{nf}$), activated macrophage, pathogenic T cells, insulin resistance, mTOR levels and beta cell antigenic protein concentrations \cite{Kodama2012}. The distinction between beta cells into functioning and non-functioning cells allows to take into account for the reduction and exhaustion of insulin produced by the beta cells. Although the preliminary outcomes of the DIRECT study suggests that beta cells can be restored to normal function through the removal of excess fat in the cells \cite{ZHYZHNEUSKAYA2019, Lean2018}, we have not taken into account the recovery of the pancreatic function. Inflammation is key in diabetes, and the interaction between inflammation and metabolism can be considered a key homeostatic mechanism \cite{Hotamisligil2006}. The model considers both the effect of exercise and dietary \cite{Schulze2005}. This model was analysed using sensitivity analysis and investigation to determine its properties (not shown). Sensitivity analyses are commonly used in inverse modelling to determine how significant each parameter is to the output variables of the system. A local analysis describes the sensitivity relative to point estimates of the parameters whereas a global analysis examines the entire parameter distribution.

\begin{equation} \label{eq:glucose}
\frac{dG}{dt}=R_{0}-
 G \Big( E_{G0}+\overbrace{S_{I}\frac{I}{I_R+i}}^{\textrm{effect of insulin resistance}} \Big) + \overbrace{R_1 H_m (t)}^{\textrm{diet}} - \overbrace{R_2 H_w (t)}^{\textrm{workouts}}  \ 
\end{equation}

\begin{equation}
\frac{dI}{dt}=\sigma \overbrace{ \frac{\beta _{f}G^{2}}{\alpha +G^{2}}}^{\textrm{effect of glucose}}-kI
\end{equation}

\begin{equation} \label{eq:beta}
\frac{d\beta_f}{dt}=-r_{0}+r_{1}G-r_{2}G^{2}\beta_f 
\end{equation}


\begin{equation} \label{eq:insulin-resistance}
\frac{dI_{R}}{dt}= -i_0 I_{R} + \overbrace{m Cyt}^{\textrm{pro-inflammatory cytokines}} + qI \ \ 
\end{equation}

where $G$ is the glucose concentration,  $I$ is Insulin concentration, $\beta_f$ functioning $\beta$-cells,  $I_{R}$ insulin resistance, and  $Cyt$ is the  concentration of pro-inflammatory cytokines \cite{Solomon2010, SC2013, Lu2014}. 
$H_m$ and $H_w$ are two step functions describing glucose intake during meals and glucose consumption during workouts respectively:

\begin{equation}
    H_m = \sum_i g_{m,i} s_{m,i} I_{t_{m,i} (1 + \Delta_m)} (t)
\end{equation}

\begin{equation}
    H_w = \sum_i z_{w,i} I_{t_{w,i} (1 + \Delta_w)} (t)
\end{equation}

where $g_m$ is the glycemic load, $s_m$ the carbohydrate serving, and $t_m$ the meal starting time; $z_w$ the number of burned calories and $t_w$ the workout starting time; $I_{t_{m,i} (1 + \Delta_m)} (t)$ and $I_{t_{w,i} (1 + \Delta_w)} (t)$ are indicator functions.

Here we consider progressive alteration of arterial stiffness and hypertension in diabetic patients. It is noteworthy that low chronic inflammation related to metabolic active abdominal obesity (abnormal secretion of adipokines and cytokines like TNF-alfa and interferon) and the impaired immune-response to infection (abnormal cytokine profile and T-cell and macrophage activation) cause an increased risk of severe COVID-19. Diabetic patients are frailer with respect to the normal population when considering COVID-19 multi-organ and multi-process disruption.




\subsection{Circulatory system model}
\label{sec:cardio}
Circulatory system models are used to describe blood flow, volume, and pressure dynamics. Equations \ref{eq:trigger1}-\ref{eq:tbv} illustrate a simplified open-loop cardiovascular model composed of five components: heart (\ref{eq:trigger1}-\ref{eq:dvlvdt}), systemic circulation (\ref{eq:pvc}-\ref{eq:dabpfoldt}), pulmonary circulation (\ref{eq:ppap}-\ref{eq:dvpvdt}), coronary circulation (\ref{eq:pcorepi}-\ref{eq:dvcorvndt}), and baroreceptors (\ref{eq:bvaso}-\ref{eq:dnidt}). Equations have been derived from the open-loop circulatory model proposed in \cite{neal2007subject}.
The heart model is composed of four sections (chambers) corresponding to right atrium, right ventricle, left atrium, and left ventricle. Each chamber is modeled as a bellows pump comprised of a one-way valve (pulmonary, tricuspid, mitral, and aortic) and a time-varying elastance (Eq. \ref{eq:eij}) controlling blood outflow \cite{rideout1991mathematical,heldt2002computational}. Blood inflow is passive. The systemic circulation has been modeled with seven vascular segments: proximal aorta, distal aorta, arteries, arterioles, capillaries, veins, and the vena cava. Each vessel has been designed using a resistance element reflecting the impact on blood flow reduction and a compliance element indicating the tendency of arteries and veins to stretch in response to pressure. High-frequency effects caused by wave reflections at great arterial bifurcations (distal and proximal aorta) are modeled with inertance elements. Arterioles, veins and vena cava have unique nonlinear PV
relationships as described in \cite{lu2001human} (see Eqs. \ref{eq:psa_a}-\ref{eq:psa}, \ref{eq:psv}, and \ref{eq:pvc}).
The pulmonary circulation is composed of five vascular segments: proximal and distal pulmonary artery, small arteries, capillaries, and veins. Wave reflections in the proximal and distal pulmonary arteries are modeled with inertance elements.
The coronary circulation model consists of four segments: epicardial and intramiocardial arteries, coronary capillaries, and coronary veins. Following \cite{neal2007subject}, large and small artery and vein segments proposed in \cite{zinemanas1994relating} have been condensed into intramiocardial arteries and coronary veins, respectively.
Baroreceptors are special sensory neurons that are excited by a stretch in the carotid sinus and aortic arch vessels. Their feedback is processed by the brain in order to maintain proper blood pressure. Baroreceptors' firing frequency to the brain has been modeled as a second-order response to the aortic pressure change \cite{di1993circulatory,lu2001human}. The second-order differential equation has been rewritten into two first order equations in order to make it compatible with common Python solvers (Eqs. \ref{eq:dnbrdt} and \ref{eq:d2nbrdt2}).

Circulatory system parameters and initial conditions have been reported in table \ref{tab:cardio}.

\subsection{Stiffness model}
\label{sec:stiff}

The complexity underlying multifactorial diseases requires the introduction of computational systems representing multi-organ and inter-process communication. To this aim, we propose a mathematical model describing the impact of comorbidities on the circulatory system. Several factors influencing blood pressure and arterial stiffness have been modeled including: diabetes, renal impairments, viral infections, lifestyle and ageing.

Ageing affects the circulatory system in multiple ways. Baroreceptors' feedback and pathways to the heart's pacemaker system decrease their efficiency over time. Heart muscle cells tend to degenerate and its walls get thicker slowing down the time the heart takes to fill with blood increasing pressure on the vessels. Additionally, blood vessels show a decrease in performance, since arteries tend to narrow and become more rigid. 

Glucose concentration affects the renin-angiotensin-aldosterone pathway as it controls the concentration and activity of Renin, ACE, and AT1R. AT1R activity is strongly related to vasoconstriction, hypertension, and inflammatory response. Hence, arterial stiffness gets even worse increasing the risks of clogged arteries and strokes. 
Besides, SARS-CoV-2 strongly bind to ACE2 decreasing its availability and impacting downstream RAS pathways regulating blood pressure. Lower levels of available ACE2 reduce the concentration of ANG-(1-7), the endogenous ligand for the G protein-coupled receptor MAS, a receptor associated with cardiac, renal, and cerebral protective responses. Hence, vasoprotection and hypotension feedbacks deteriorate increasing inflammatory response and pressure on blood vessels.


The combined effect of comorbidities and ageing factors on arterial stiffness and inflammation may lead to critical circulatory conditions and fibrosis. High glucose concentrations strengthen RAS hypertension feedbacks and lower blood vessels' lumen, especially on capillaries, arterioles, and venules. By affecting blood pressure regulation pathways, SARS-CoV-2 infections may impair vasoprotection regulation by the RAS endangering the whole circulatory system with disruptive repercussions among the elderly. The combination of all such factors may lead to acute diseases such as thrombophlebitis, cardiomyopathy, myocardial infarction, pulmonary embolism, heart failure, and eventually to patients' death.

The diabetic model (Sec. \ref{sec:diabetes}) accounts both for hyperglycemic conditions and lifestyle habits. After lunch and dinner, glucose concentration in blood vessels peaks, while it is scaled down by insulin or physical exercise.
The RAS model (Sec. \ref{sec:ras}) has been used to simulate peptides' and drug concentration dynamics taking into account glucose concentration, ACE inhibitor treatments, renal conditions, and viral infections binding to ACE2 (such as COVID-19).
Abnormal ACE2 activity ($k_{ACE2} - k_{ACE2,0}$) has been assumed as proportional to SARS-CoV-2 infectivity (see Eq. \ref{eq:ace2}). ACEi or ARB treatments could also increase ACE2 abundance and thus enhance viral entry \cite{south2020controversies}. In case of severe renal conditions, only a fraction of drug diacid is expelled before the subsequent administration (see Eq. \ref{eq:pk} and Fig. \ref{fig:RD}). The drug surplus left inside the body may reinforce inflammation.
Overall, the inflammatory response has been modeled as a function of all such contributions:

\begin{eqnarray} \label{eq:is}
    \frac{d\text{IR}}{dt} &=& \overbrace{k_{SARS} (k_{ACE2} - k_{ACE2,0})}^{\textrm{viral infection}} + \overbrace{k_D [Drug]}^{\textrm{drug treatment}} + \overbrace{k_G G}^{\textrm{glucose}} - \overbrace{k_{eff} \text{IR}}^{\textrm{anti-inflammatory response}}
\end{eqnarray}\todo{heparin?\\ https://www.ncbi.nlm.nih.gov/pmc/articles/PMC1903448/\\
https://onlinelibrary.wiley.com/doi/epdf/10.1111/jth.14817}

where $k_{SARS}$ represents SARS-CoV-2 affinity with ACE2, $k_D$ the inflammation rate due to ACEi surplus, $k_G$ the inflammation rate due to glucose surplus, and $k_{eff}$ the anti-inflammatory response rate.

One of the main processes associated with arterial stiffness occurring during ageing is DNA methylation, consisting in the addition of methyl groups to the DNA molecule which may modify the activity of a DNA segment without changing the sequence. DNA methylation has been modeled as a linear function of the age $A$ \cite{Wen2015}:

\begin{equation}\label{eq:met}
    \alpha_{MET} = \beta_0 - \beta_1 A \qquad \beta_{0,1} \geq 0
\end{equation}

As a result, blood vessels' compliance parameters have been reduced by a factor accounting for the combined effect of inflammation (Eq. \ref{eq:is}) and ageing (Eq. \ref{eq:met}):

\begin{equation}
    \widehat{C}_i = \overbrace{\alpha_{MET} \Big(1 - \frac{\text{IR}}{100} \Big)}^{\textrm{stiffness}} \ C_i
\end{equation}

where $C_i$ is the compliance of the blood vessel $i$ for a young healthy individual and $\widehat{C}_i$ is the reduced compliance.
The circulatory model (Sec. \ref{sec:cardio}) has been used to simulate blood pressure dynamics in critical vessels where blood pressure spikes may lead to acute diseases.

\subsection{Extending the model to COVID-19 treatments}
One of the main issues related to COVID-19 is blood clotting.
Studies have reported that 30\% of COVID-19 patients showed signs of blood clots in their lungs.
One of the recommendations is to give a low dose of heparin, which prevents clot formation or tissue plasminogen activator (tPA), which helps to dissolve blood clots \cite{Sardu2020,Tang2020}.
Besides, several observational studies and clinical trials reported that vitamin D supplementation reduced the risk of influenza and inflammation by raising its blood concentrations above 40-60 ng/mL (100-150 nmol/L) \cite{lisse2011vitamin,pilz2009vitamin,cohen2006vitamin,grant2020evidence}.
Hence, we extended our model by taking into account such preliminary COVID-19 treatments. In fact, both heparin and vitamin D have an indirect impact on blood pressure by making blood less dense, reducing clotting formation, and lowering inflammation. We modeled the impact of such treatments by including additional terms on blood pressure inside the cardiovascular model:

\begin{equation}
    \widehat{p_i} = p_i - \beta_h [heparin] - \beta_D ([D] - [D]_0)
\end{equation}

where $[heparin]$ is the heparin dosage and $[D]$ the vitamin D concentration.

\section{Experiments}
\label{sec:exp}

\begin{figure}
    \centering
    \includegraphics[scale=0.6]{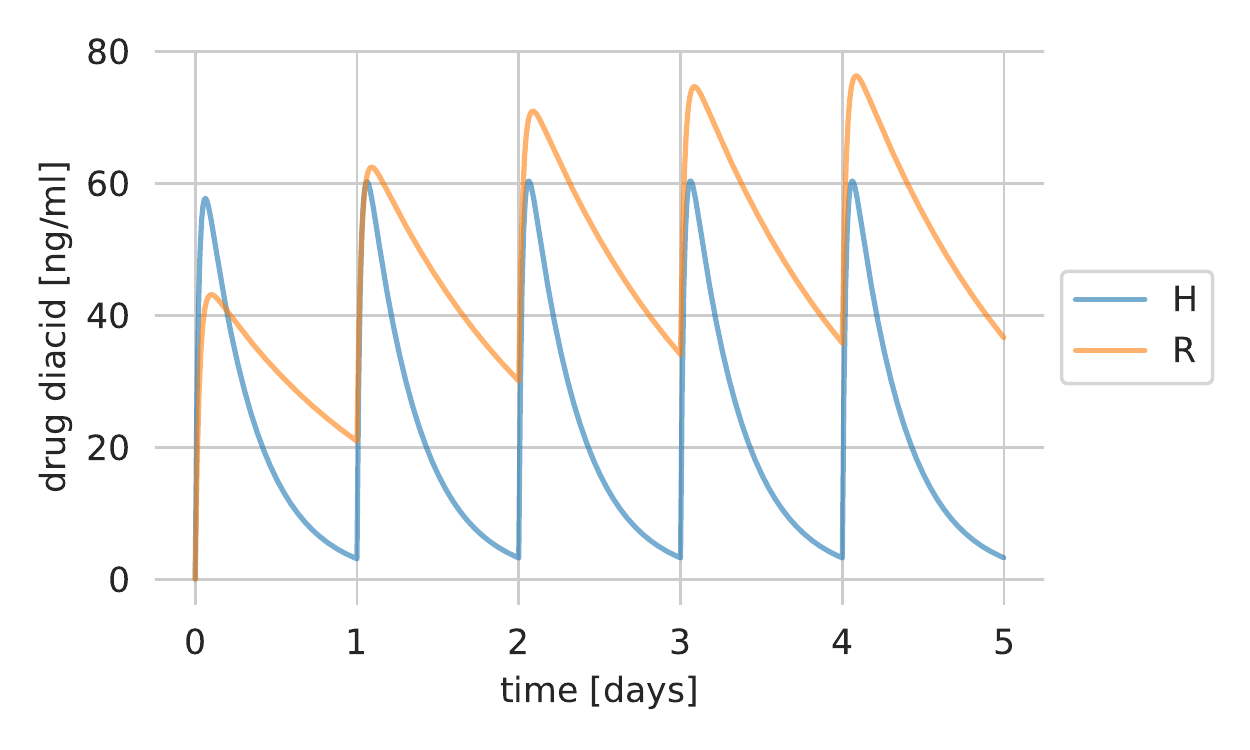}\qquad
    \includegraphics[scale=0.6]{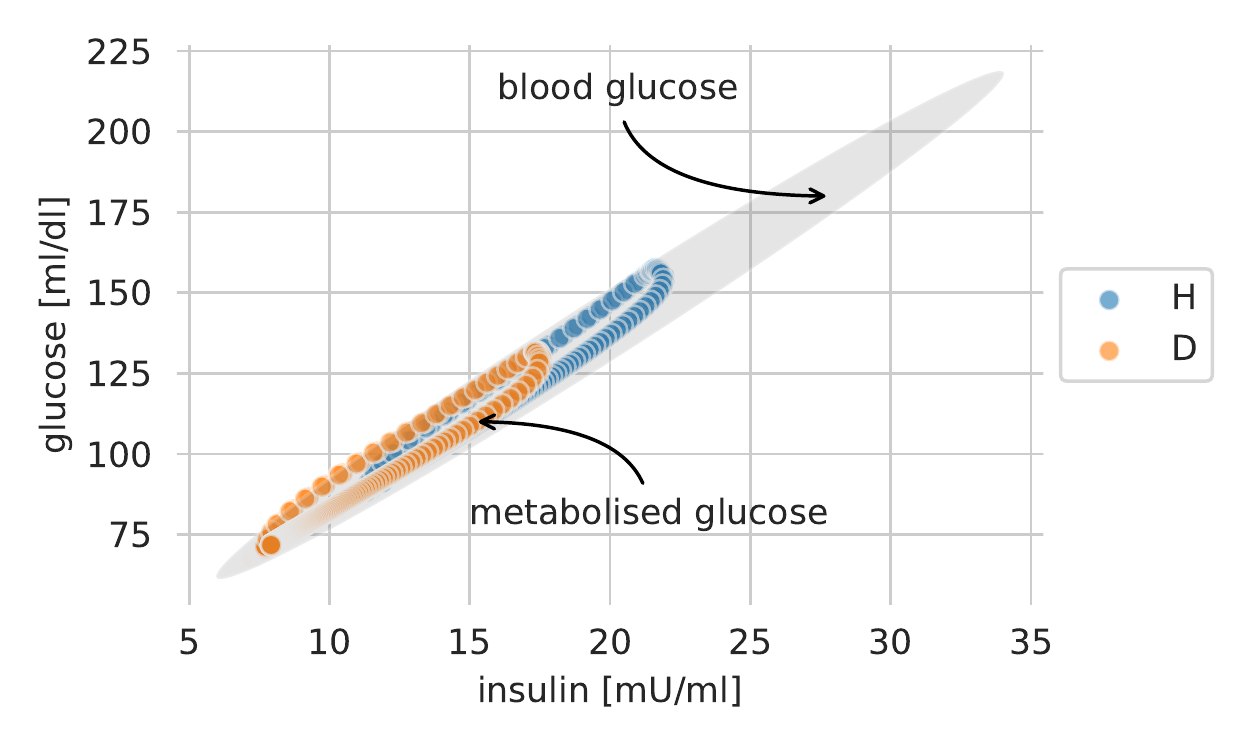}
    \caption{Drug concentration for healthy individuals and patients with renal impairments (\textbf{left}). Glucose metabolised by insulin in healthy and diabetic individuals (\textbf{right}).}
    \label{fig:RD}
\end{figure}

\begin{figure}[!ht]
    \centering
    \includegraphics[scale=0.6]{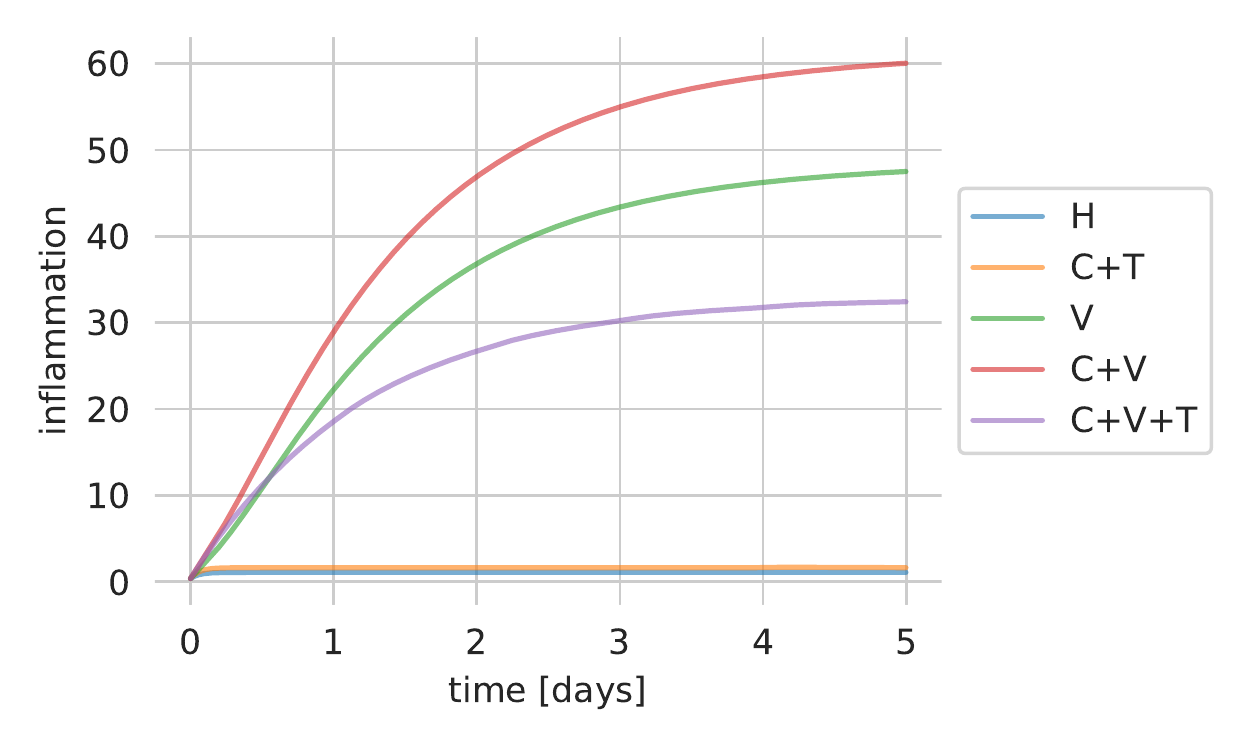}\qquad
    \includegraphics[scale=0.6]{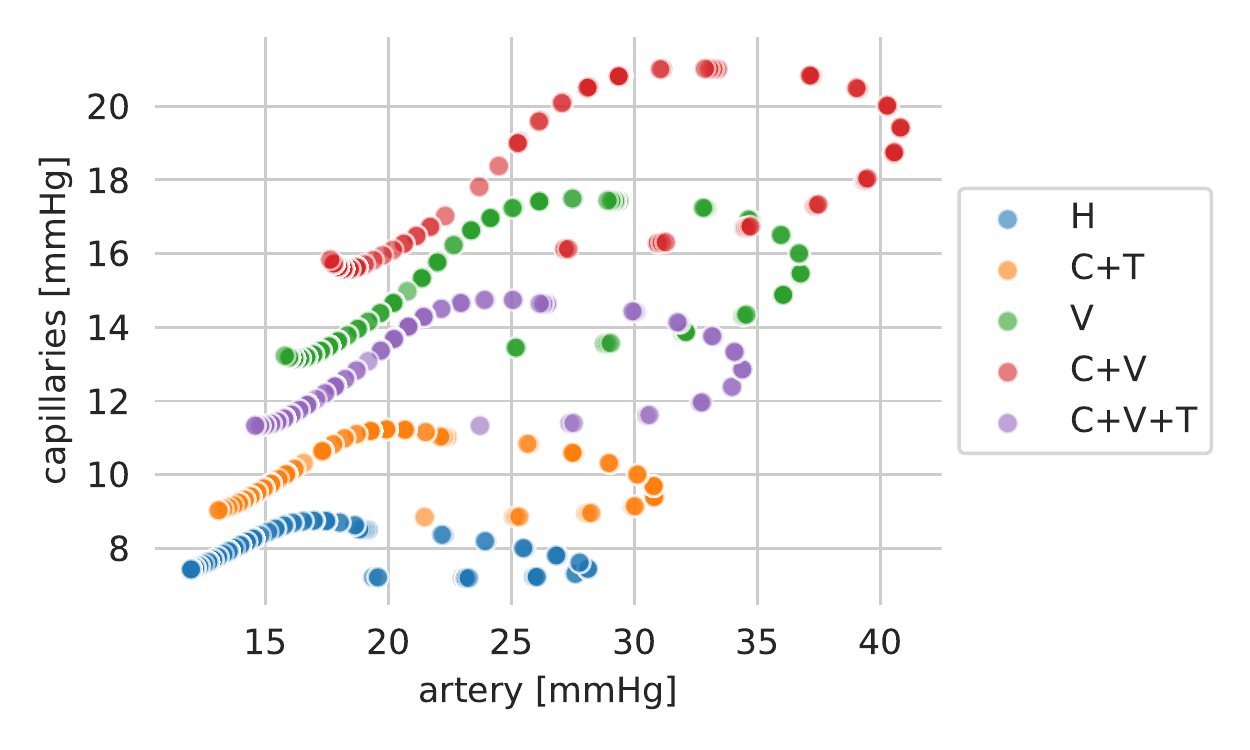}\\
    \includegraphics[scale=0.6]{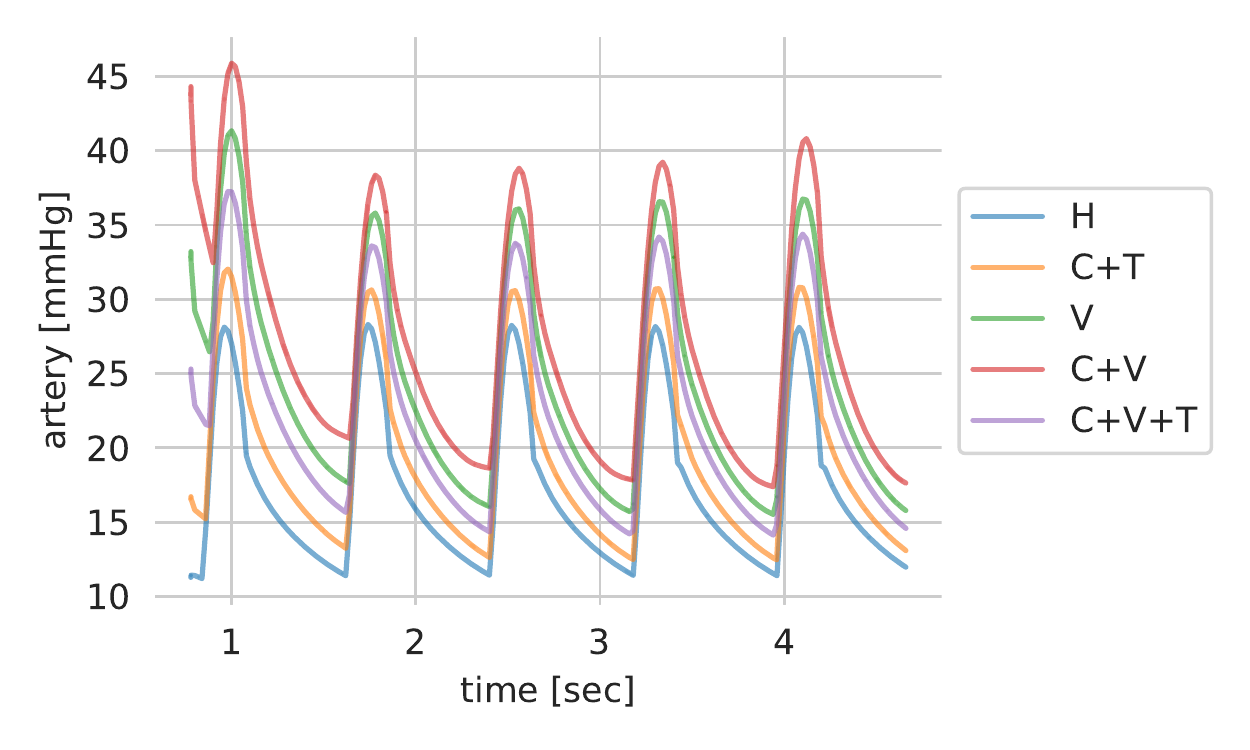}\qquad
    \includegraphics[scale=0.6]{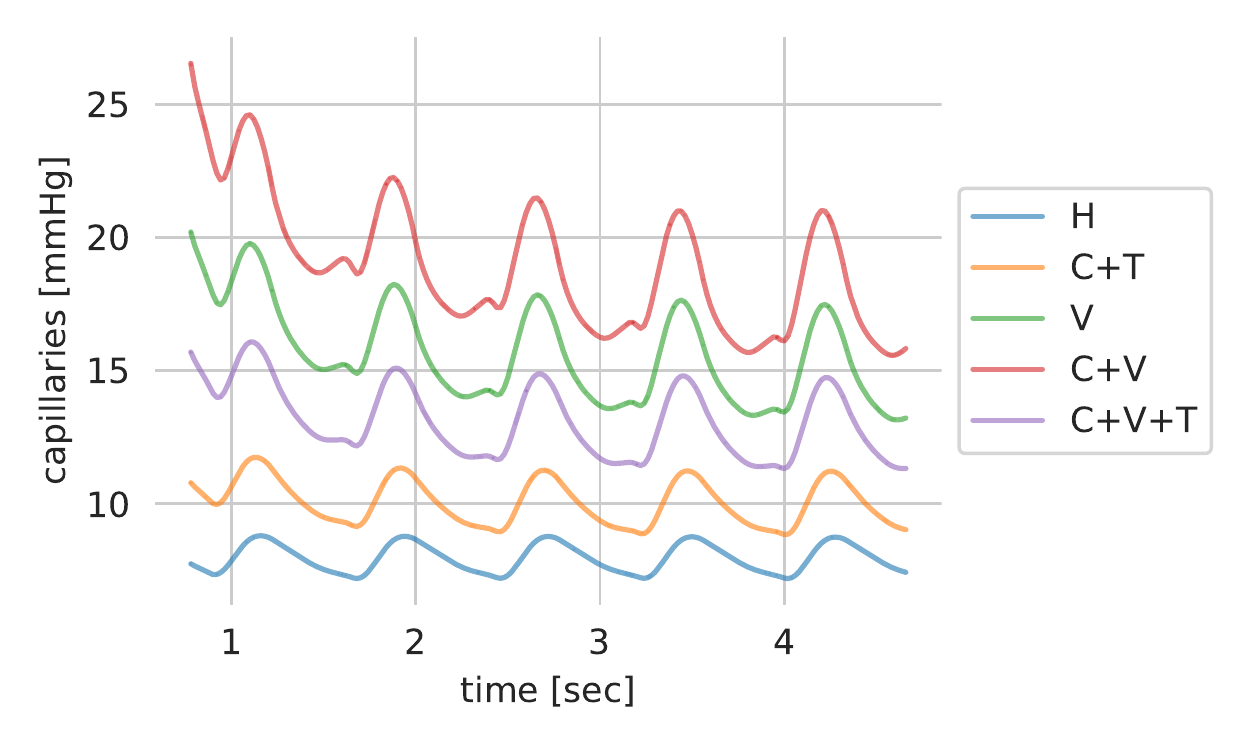}\\
    \caption{Inflammation scores (\textbf{top-left}, see Eq. \ref{eq:is}), the corresponding lungs' pressures phase space (\textbf{top-right}) and dynamics over time (\textbf{bottom})}
    \label{fig:lungs}
\end{figure}

The models presented in Section \ref{sec:methods} have been solved to analyze the effects of comorbidities like diabetes, renal impairment, and viral infections affecting the circulatory system. Table \ref{tab:experiments} reports the set of experimental conditions that have been analysed. Five computational patients have been created corresponding to different physiological states. These scenarios have been further stratified by the age of the computational patient, given that arterial stiffness has been modeled as a function of the increased DNA methylation during ageing. Drug concentrations (Fig. \ref{fig:RD}), inflammation levels, and blood pressure dynamics in lung vessels (Fig. \ref{fig:lungs}) in comorbid conditions have been compared to the dynamics obtained in healthy states or using ACE inhibitor treatments.

\begin{table}[!ht]
\renewcommand{\arraystretch}{1.5}
\centering
\caption{computational patients' conditions used for the simulations. The diabetic and the RAS models do not depend on patient's age. Lifestyle habits have been set as three meals and one light workout session in the afternoon for all patients.}
\label{tab:experiments}
\begin{tabular}{lll}
\hline
\textbf{Label} & \textbf{Age} & \textbf{Description} \\ 
\hline
H & 20 & Healthy individual \\
D & - & Diabetic individual \\
R & - & Individual with renal impairment \\
C+T & 70 & Individual with comorbid conditions (diabetes + renal impairment) treated with ACEi \\
V & 70 & Individual with COVID-19 \\
C+V & 70 & Individual with comorbid conditions and COVID-19 \\
C+V+T & 70 & Individual with comorbid conditions and COVID-19 treated with ACEi \\
C+V+3T & 70 & Individual with comorbidities and COVID-19 treated with ACEi, heparin, and vitamin D \\
\hline
\end{tabular}
\end{table}

The RAS model has been simulated for constant glucose cases using the daily glucose peak predicted by the diabetic model right after the main meals. Glucose concentration ranged between the extremes of normal glucose at $6-7$ $mmol/L$ (corresponding to $108-125$ $ml/dl$) and high glucose at $10-11$ $mmol/L$ (corresponding to $180-200$ $ml/dl$) based on experimental studies \cite{durvasula2008activation,yard2002influence,das2014high}. The time window of the RAS simulations has been set to five days, corresponding to five daily ACEi administrations \cite{zaman2002drugs}. The simulation results have been used to compute arterial stiffness and to reduce compliance parameters of blood vessels in the open-loop circulatory model. In the following simulations the arterial blood pressure (ABP) signal used in \cite{neal2007subject} has been used instead of personalised clinical measurements.

Fig. \ref{fig:RD} shows the dynamics of the concentration of ACEi and glucose-insulin dynamics over the first five days of treatment. Due to renal impairment, the computational patient was not able to expel the drug dose before the next administration. The inflammatory response and the corresponding blood pressure dynamics in lungs' vessels are shown in Fig. \ref{fig:lungs}. Comorbid conditions tend to increase blood pressure variability in all scenarios. However, as arterial stiffness grows with the age of the computational patient, the variability increases as well, possibly leading to irreversible deterioration of blood vessels' walls. ACEi treatments may help in reducing inflammation levels, but may not be sufficient to recover healthy blood pressures.
One of the most serious effects illustrated by simulations consists of an increased  mean value of blood pressure and blood pressure variability especially on small pulmonary vessels and capillaries (see Fig. \ref{fig:lungs}), increasing the risk of clogged arteries, fibrosis, and strokes. Besides, experimental results shows how fluctuations of variables over time may change and present different shapes especially on small vessels. In computational patients with comorbidities blood pressure dynamics in pulmonary capillaries exhibit higher mean values and variability, but beat frequencies can be observed as well.

\subsection{COVID-19}
The COVID-19 mortality statistics underline the relevance of deeper analysis on multi-factorial diseases in fighting the pandemic \cite{Cariou2020}. Underlying morbidities such as cardiovascular disease, kidney disease, T2D or tumours have been observed in patients with severe infection, especially among the elderly \cite{Wang2020}. By affecting blood pressure regulation pathways, SARS-CoV-2 infections may impair vasoprotection regulation endangering the whole circulatory system with severe repercussions. By taking advantage of our composable framework, experimental results offer an overview on how the combination of multiple diseases with SARS-CoV-2 may lead to acute conditions. Fig. \ref{fig:lungs} clearly shows how the computational patient with comorbidities and SARS-CoV-2 has higher risk of pulmonary vessels' deterioration.

The combined effect of heparin and vitamin D can help in reducing blood pressure mean by making the blood more fluid. However, they do not affect blood pressure variability determined by vessels' rigidity. Hence, the risk in developing cardiovascular diseases related to blood pressure variability may still be high despite treatments.
Notably the results of our experiments agree with hypotheses suggesting that healthy blood vessels protect children from serious effects of COVID-19.

It is noteworthy that autopsy-based findings have demonstrated a variety of damages caused by COVID-19 infection, among which extensive coagulopathy, acquired thrombophilia and endothelial cell death \cite{Becker2020}. Here we consider the sole effect on blood pressure.

\begin{figure}[!ht]
    \centering
    \includegraphics[scale=0.8]{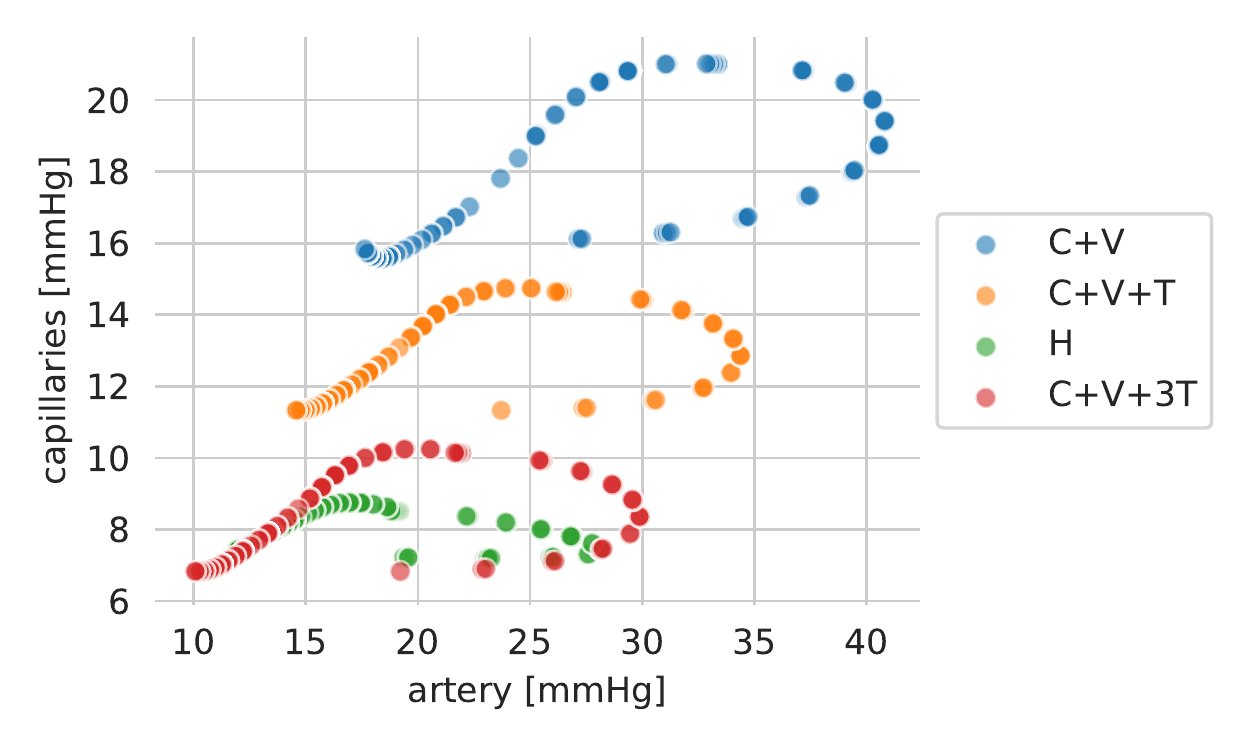}
    \caption{Comparing blood pressure effects of COVID-19 treatments in pulmonary vessels.}
    \label{fig:heparin}
\end{figure}

\section{Discussion}

The modularity and composability of different available mechanistic and phenomenological models presents the challenge to define a mathematical framework connecting different systems’ descriptions, their dynamics, and constraints. Let's imagine to put together a model based on ODEs and a model in terms of a discrete space discrete time Markov chain. This has then to be done in the light of behavioral properties that can be sets of trajectories or measures on the trajectory space (typically those learned from data with statistical methods). Cell-level models (using ODEs, delay differential equations, DDE, or agents) need to be systematically scaled up to the tissue level; for the multiple timescale problems, the challenge is to obtain a model order reduction, i.e. to abandon high dimensional bioengineering systems in favour of simpler effective mathematical models. The tissue level could be modeled using PDE or cellular Potts model which may provide better representation for detailed and heterogeneous cell-cell,
cell-tissue, cell-matrix interaction cases. Integrative models, could be made by single scale models, describing the biological process at different characteristic space-time scales, and scale bridging models, which define how the component models are coupled to each other. While at the tissue level, physical quantities usually vary across space and time, in a continuous fashion, and can be thus represented using systems
of PDE \cite{Viceconti2011}.

\subsection{Emerging properties of variances from model composability }

Many physiological variables have a circadian trend; sometimes also a seasonal one. For example blood pressure decreases during sleep and shows a sharp uprise at the time of awakening. This early morning variation is often concurrent with an increase of acute myocardial infarction, sudden cardiac death, and stroke
\cite{ELLIOTT1999}. 
Common clinical parameters such as diastolic and systolic blood pressure, heart patterns, blood cell counts are usually evaluated as averages. Little importance is given to higher moments such as variances during the day or during a longer interval of time. The lack of continuous measures for most of the quantities has generated a medical practice that disregard of unobserved or partially observed data. Some authors identified a disease and age-related loss of complex variability in multiple physiologic processes including cardiovascular control, pulsatile hormone release, and ECG data \cite{Lipsitz1992,Glass1989}.
Our composable model reveals interesting patterns, particularly fluctuations in blood pressure, particularly during COVID acute infection when the diabetic model is coupled with the RAS and the cardiovascular models. We believe that the use of extensive models could enable to understand concurrent patterns of alteration in different districts.

\subsection{How such computational patient model could be deployed and further developed}

computational Patient will benefit from using machine learning and data analysis of large amount of data such for instance that obtained from UK Biobank as modeling will have a truly catalytic effect in synergy with machine learning. 
The computational patient model requires adequate artificial intelligence support
to generate diagnosis and validate its correctness. A decision-making process
could be based on the development of a personalised statistics of changes in health, end-stage disease, signs and symptoms (CHESS, \cite{Hirdes2003}).
This ideally would develop through monitoring of the individualized response to 
therapeutic interventions, in addition to changes in risk profile.
One aspect is a dedicated CHESS scale based on all the variables and observable considered considered by the model(s) \cite{Hirdes2003,Ades2008}.
It will act as a personalised patient simulator and will draw
temporal trajectories of disease and comorbidities progression. The
trajectories will change with drug regimes, medical intervention, and
lifestyle changes.

Any data used will be anonymises or de-identified using ad hoc software (see for instance \cite{Cardinal2017}) and  we will follow the FAIR principle (Findable, Accessible, Interoperable, Reusable) and the GRPD regulation. One meaningful approach to extract useful indication is to use a clinical decision support system which incorporates medical experience, research results and personal judgement \cite{Mller2020}. We believe the computational patient models to be in a research only state and therefore we do not make further integrations.

The future foreseen is that AI will assist our health and disease conditions in a more effective way than nowadays: a medical check up will be supported by well-tuned artificial intelligence and patient-based modeling . At the clinical level, computer-aided therapies and treatments will develop into intervention strategies undertaken under acute disease conditions or due to external factors (infections) to contrast cascade effects. In non acute states, predictive inference will propose prevention plans for comorbidity management, particularly in presence of multiple therapies.

Therefore this approach is meaningful in perspective of a computational medicine characterised by a close coupling between bioinformatics, clinical measures
and modeling prediction and perhaps remote patient monitoring.

\section{Conclusion}
computational scientists and bioengineers' vision is a framework of methods and technologies that, once established, will make it possible to investigate the human body as a whole. It calls for a total transformation in the way healthcare currently works and is delivered to patients.
Underpinning this transformation is substantial technological innovation with a requirement for deeper trans-disciplinary research, improved IT infrastructure, better communication, large volumes of high quality data and machine learning and modeling tools. Machine learning could be automatised (i.e. autoML) and models should be modular so to be organised to answer specific and personalised medical questions.
Simulations are increasingly regarded as valuable tools in a number of aspects of medical practice including lifestyle changes, surgical planning  and medical interventions.  The idea is that cross-modality data is obtained for the patient and machine learning techniques estimate parameters to be input into modeling framework.
We believe that a deeper understanding and practice of modeling in medicine
will produce better investigation of complex biological processes, and even new ideas and better feedback into medicine.
Finally, computational models are cheap and this will make possible to predict drugs interaction and to make better use of generic drugs. In this sense the personalised model will become a product associated with the drug.

\subsection{Disclaimer}

\textbf{The computational tool has not been validated and should not be used for clinical purposes.}\\
To enable code reuse, the Python code for the mathematical models including parameter values and documentation is freely available under Apache 2.0 Public License from a GitHub repository \cite{repo}.
Unless required by applicable law or agreed to in writing, software is distributed on an "as is" basis, without
warranties or conditions of any kind, either express or implied.

\section{Acknowledgement}
The authors have received funding from the European Union’s Horizon 2020 research and innovation programme under grant agreement No 848077. We thank Gianluca Ascolani, Rachel Clark, Annalisa Occhipinti, Stefan Stojanovic, Emma Rocheteau, Jacob Deasy, Alberto Giannoni, Alberto Tonda, and Elisabetta Zara for suggestions.

\bibliographystyle{unsrt}  
\bibliography{references}  

\clearpage
\appendix
\section{Appendix}

\subsection{Website}

\begin{figure}[!ht]
    \centering
    \includegraphics[scale=0.3]{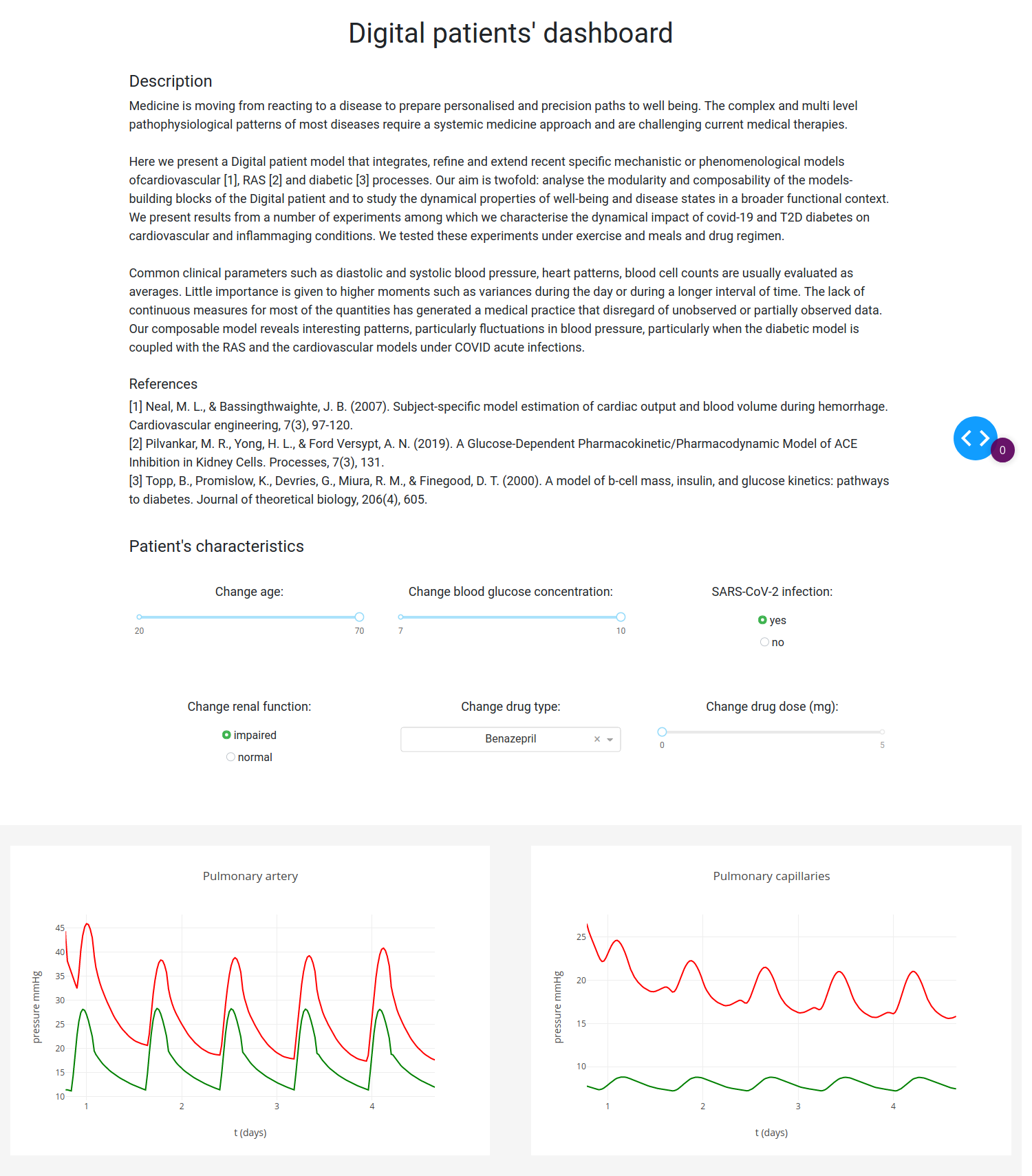}
    \caption{Website dashboard for computational patients.}
    \label{fig:website}
\end{figure}
\clearpage  












\subsection{Equations of the renin-angiotensin system}

\begin{equation} \label{eq:drug}
    I = \frac{100 [DRUG]^m}{[DRUG]_{50}^m + [DRUG]^m}
\end{equation}

\begin{eqnarray}
    \frac{d[AGT]}{dt} &=& 
    \overbrace{k_{AGT}}^{\textrm{production rate}} - 
    \overbrace{\Big( a_{Renin} G + b_{Renin} \Big) [AGT]}^{\textrm{renin-catalyzed conversion to ANG I}} - 
    \overbrace{\frac{ln(2)}{h_{AGT}} [AGT]}^{\textrm{degradation}}
\end{eqnarray}

\begin{eqnarray} \label{eq:renin}
    \frac{d[Renin]}{dt} &=& 
    \overbrace{\frac{\ln{2}}{h_{Renin}} [Renin]_0}^{\textrm{production rate}} \nonumber \\ 
    &+& \overbrace{\frac{k_{f,sys} [ANGII]_{0,sys}}{[ANGII]_0} \Big( [ANGII]_0 - [ANGII] \Big) \Bigg(1 - ([ANGII]_0 - [ANGII]) \frac{[ANGII]_{0,sys}}{f_{sys} [ANGII]_0} \Bigg)}^{\textrm{ANG II inhibition feedback}} \nonumber \\
    &-& \overbrace{\frac{\ln{2}}{h_{Renin}} [Renin]}^{\textrm{degradation}}
\end{eqnarray}

\begin{eqnarray}
    \frac{d[ANG I]}{dt} &=& 
    \overbrace{\Big( a_{Renin} G + b_{Renin} \Big) [AGT]}^{\textrm{renin-catalyzed conversion of AGT}} + 
    \overbrace{k_{Renin} \Big( [Renin] - [Renin]_0 \Big)}^{\textrm{ANG II feedback on renin}} \nonumber \\
    &-& \overbrace{\Big( a_{ACE} G + b_{ACE} \Big) [ANG I] \Big(1 - I\Big)}^{\textrm{ACE-catalyzed conversion to ANG II subject to inhibition}} 
    - \overbrace{\Big( k_{NEP} + k_{ACE2} \Big) [ANG I]}^{\textrm{conversion to ANG-(1-7) and ANG-(1-9)}} - \nonumber \\
    &-& \overbrace{\frac{ln{2}}{h_{ANG I}} [ANG I]}^{\textrm{degradation}}
\end{eqnarray}

\begin{eqnarray} \label{eq:ANG2}
    \frac{d[ANG II]}{dt} &=& 
    \overbrace{\Big( a_{ACE} G + b_{ACE} \Big) [ANG I] \Big( 1 - I \Big)}^{\textrm{ACE-catalyzed conversion to ANG II subject to inhibition}} \nonumber \\
    &-& \overbrace{\Big( k_{ACE2} + \Big( a_{AT1} G + b_{AT1} \Big) + k_{AT2} + k_{APA} \Big) [ANG II]}^{\textrm{conversion to AT1R, AT2R, APA, and ANG-(1-7)}} \nonumber \\
    &-& \overbrace{\frac{ln{2}}{h_{ANG II}} [ANG II]}^{\textrm{degradation}}
\end{eqnarray}

\subsection{Equations of the open-loop circulatory model}

\subsubsection{Four-chambered heart}

\begin{equation} \label{eq:trigger1}
    t \geq t_{HB}(n+1) - PR_{int} - offv \implies
        \begin{dcases}
            HR_a = \frac{1}{HP / 60} \\
            T_{s_a} = Ts1_a \sqrt{ \frac{Ts2}{HR_a / 60} } \\
            t_{Pwave} = t_{HB} - PR_{int} - offv \\
            n = n + 1
        \end{dcases}
\end{equation}

\begin{equation}
    t \geq t_{HB}(m+1) - offv \implies
        \begin{dcases}
            HR_v = \frac{1}{HP / 60} \\
            T_{s_v} = Ts1_v \sqrt{ \frac{Ts2}{HR_v / 60} } \\
            t_{Rwave} = t_{HB} - offv \\
            m = n \\
            V_{var,i,vs0} = \begin{dcases} 
            V_{i,vd0} &\qquad V_{i,v} < V_{i,vd0} \\
            V_{i,vs0} &\qquad V_{i,v} > EDV_{i_v} \\
            V_{i,vs0} - V_{i,vd0} \frac{V_{i,v} - V_{i,vd0}}{EDV_{i,v} - V_{i,vd0}} + V_{i,vd0} &\qquad V_{i,vd0} \leq V_{i,v} \leq EDV_{i_v}
            \end{dcases}\\
            af_{con2} = af_{con}
        \end{dcases}
\end{equation}

\begin{equation}
    HR = HR_v
\end{equation}

\begin{equation} 
    E_{max,i,v} = K_{e,i,v} E_{max,i,v1}
\end{equation}

\begin{equation}
    t_{a,REL} = t - t_{Pwave}
\end{equation}

\begin{equation}
    t_{v,REL} = t - t_{Rwave}
\end{equation}

\begin{equation} 
    y_i = 
    \begin{dcases}
    \dfrac{1 - \cos \pi \dfrac{t_{i,REL}}{T_{s,i}} }{2} &\qquad 0 \leq t_{i,REL} < T_{s,i} \\
    \dfrac{1 + \cos 2 \pi \dfrac{t_{i,REL} - T_{s,i}}{T_{s,i}} }{2} &\qquad T_{s,i} \leq t_{i,REL} < 1.5 T_{s,i} \\
    0 &\qquad t_{i,REL} \geq 1.5 T_{s,i}
    \end{dcases}
\end{equation}

\begin{equation} \label{eq:eij} 
    E_{i,j} = y_j (E_{max,i,j} - E_{min,i,j}) + E_{min,i,j}
\end{equation}

\begin{equation} 
    V_{i,a,0} = (1 - y_a) (V_{i,a,d0} - V_{i,a,s0}) + V_{i,a,s0}
\end{equation}

\begin{equation} 
    V_{i,v,0} = (1 - y_v) (V_{i,v,d0} - V_{var,i,vs0}) + V_{var,i,vs0}
\end{equation}

\begin{equation}
    \psi(v) = K_{xp} \frac{1}{e^{v / K_{xv}} - 1}
\end{equation}

\begin{equation} 
    P_{i,a} = E_{i,a} (V_{i,a} - V_{i,a0}) - \psi(V_{i,a})
\end{equation}

\begin{equation} 
    P_{i,v} = E_{i,v} (V_{i,v} - V_{i,v0}) af_{con2} - \psi(V_{i,v})
\end{equation}

\begin{equation} 
    F_{i,a} = 
    \begin{dcases}
    \frac{P_{i,a} - P_{i,v}}{R_{i,a}} &\qquad P_{i,a} > P_{i,v} \\
    0 &\qquad P_{i,a} \leq P_{i,v}
    \end{dcases}
\end{equation}

\begin{equation}
    F_{r,v} = 
    \begin{dcases}
    \frac{P_{r,v} - P_{pap}}{R_{r,v}} &\qquad P_{r,v} > P_{pap} \\
    0 &\qquad P_{r,v} \leq P_{pap}
    \end{dcases}
\end{equation}

\begin{equation}
    F_{l,v} = 
    \begin{dcases}
    \frac{P_{l,v} - P_{aop}}{R_{l,v}} &\qquad P_{l,v} > P_{aop} \\
    0 &\qquad P_{l,v} \leq P_{aop}
    \end{dcases}
\end{equation}

\begin{eqnarray}
    \frac{dV_{r,a}}{dt} &=& 
\end{eqnarray}

\begin{equation}
    \frac{dV_{r,a}}{dt} = \frac{F_{vc} + F_{corvn}}{F_{r,a}}
\end{equation}

\begin{equation}
    \frac{dV_{r,v}}{dt} = \frac{F_{r,a}}{F_{rv}}
\end{equation}

\begin{equation}
    \frac{dV_{l,a}}{dt} = \frac{F_{pv}}{F_{l,a}}
\end{equation}

\begin{equation} \label{eq:dvlvdt}
    \frac{dV_{l,v}}{dt} = \frac{F_{l,a}}{F_{l,v}}
\end{equation}

\subsubsection{Systemic circulation}

\begin{equation} \label{eq:pvc}
    P_{vc} = 
    \begin{dcases}
    K_1 (V_{vc} - V_{vc0}) - \psi (V_{vc}) &\qquad V_{vc} > V_{vc0} \\
    D_2 + K_2 e^{V_{vc} / V_{min,vc}} - \psi (V_{vc}) &\qquad V_{vc} \leq V_{vc0} \\
    \end{dcases}
\end{equation}

\begin{equation}
    CO_{mod} = F_{rv,sm}
\end{equation}

\begin{equation}
    SV = \frac{CO_{mod}}{HR}
\end{equation}

\begin{equation} 
    ABP_{shift} = ABP_{meas} (t - offv)
\end{equation}

\begin{equation}
    K_v = K_{v1} K_{sv}
\end{equation}

\begin{equation}
    MAP_{mod} = \dfrac{R_{crb} \Bigg[ R_{taod} AOF_{mod} - R_{taod} F_{aod} + \dfrac{V_{aod} - V_{aod0}}{C_{aod}} - \psi (V_{aod}) \Bigg] + P_{vc} R_{taod}}{R_{crb} + R_{taod}}
\end{equation}

\begin{equation}
    P_{aod} = ABP_{shift}
\end{equation}

\begin{equation}
    P_{sap} = \frac{V_{sap} - V_{sap0}}{C_{sap}} - \psi (V_{sap})
\end{equation}

\begin{equation} \label{eq:psa_a}
    P_{sa,a} = K_c \log_{10} \Bigg[ \frac{V_{sa} - V_{sa0}}{D_o} + 1 \Bigg]
\end{equation}

\begin{equation}
    P_{sa,p} = K_{p1} e^{\tau_p (V_{sa} - V_{sa0})} + K_{p2} (V_{sa} - V_{sa0})^2
\end{equation}

\begin{equation} \label{eq:psa}
    P_{sa} = f_{vaso} P_{sa,a} + (1 - f_{vaso}) P_{sa,p}
\end{equation}

\begin{equation}
    P_{sc} = \frac{V_{sc} - V{sc0}}{C_{sc}}  - \psi (V_{sc})
\end{equation}

\begin{equation} \label{eq:psv}
    P_{sv} = -K_v \log_{10} \Bigg[ \frac{V_{max,sv}}{V_{sv}} - 0.99 \Bigg]
\end{equation}


\begin{equation}
    R_{sa} = K_r \Bigg( e^{4 f_{vaso}} + \frac{V_{sa,max}^2}{V_{sa}^2} \Bigg) + R_{sa0}
\end{equation}

\begin{equation}
    R_{vc} = K_R \frac{V_{max,vc}^2}{V_{vc}^2} + R_0
\end{equation}

\begin{equation}
    F_{crb} = \frac{MAP_{mod} - P_{vc}}{R_{crb}}
\end{equation}

\begin{equation}
    F_{sap} = \frac{P_{sap} - P_{sa}}{R_{sap}}
\end{equation}

\begin{equation}
    F_{sa} = \frac{P_{sa} - P_{sc}}{R_{sa}}
\end{equation}

\begin{equation}
    F_{sc} = \frac{P_{sc} - P_{sv}}{R_{sc}}
\end{equation}

\begin{equation}
    F_{sv} = \frac{P_{sv} - P_{vc}}{R_{sv}}
\end{equation}

\begin{equation}
    F_{vc} = \frac{P_{vc} - P_{ra}}{R_{vc}}
\end{equation}

\begin{equation}
    \frac{dV_{aop}}{dt} = \dfrac{P_{aop} - \dfrac{V_{aop} - V_{aop0}}{C_{aop}}}{R_{taop}}
\end{equation}

\begin{equation}
    \frac{dV_{aod}}{dt} = AOF_{mod} - F_{aod} - F_{crb}
\end{equation}

\begin{equation}
    \frac{dV_{sa}}{dt} = F_{sap} - F_{sa}
\end{equation}

\begin{equation}
    \frac{dV_{sap}}{dt} = F_{aod} - F_{sap}
\end{equation}

\begin{equation}
    \frac{dV_{sc}}{dt} = F_{sa} - F_{sc}
\end{equation}

\begin{equation}
    \frac{dV_{sv}}{dt} = F_{sc} - F_{sv}
\end{equation}

\begin{equation}
    \frac{dV_{vc}}{dt} = F_{sv} + F_{crb} - F_{vc}
\end{equation}

\begin{equation}
    \frac{dAOF_{mod}}{dt} = (MAP_{meas} - MAP_{mod}) \frac{K_{comap}}{60}
\end{equation}

\begin{equation}
    \frac{dF_{rv,sm}}{dt} = \frac{F_{rv} - F_{rv,sm}}{\tau_{co}}
\end{equation}

\begin{equation}
    \frac{dF_{aod}}{dt} = \frac{MAP_{mod} - F_{aod} R_{aod} - P_{sap} }{L_{aod}}
\end{equation}

\begin{equation}
    \frac{dP_{paop}}{dt} = \dfrac{ F_{lv} - \dfrac{dV_{aop}}{dt} - F_{aop} - F_{corepi} }{C_{corepi}}
\end{equation}

\begin{equation}
    \frac{dMAP_{meas}}{dt} = \frac{ABP_{shift} - MAP_{meas}}{\tau_{MAP}}
\end{equation}

\begin{equation}
    \frac{dCO_{mea}}{dt} = \frac{PAF_{meas} - CO_{mea}}{\tau_{co}}
\end{equation}

\begin{equation} \label{eq:dabpfoldt}
    \frac{dABP_{fol}}{dt} = \frac{ABP_{shift} - ABP_{fol}}{\tau_{ABP}}
\end{equation}

\subsubsection{Pulmonary circulation}

\begin{equation} \label{eq:ppap}
    P_{pap} = 
    \begin{dcases}
    P_{pap1} = \dfrac{R_{tpap} P_{rv} - R_{rv} F_{pap} R_{tpap} + \Bigg( R_{rv} \dfrac{V_{pap} - V_{pap0}}{C_{pap}} - \psi (V_{pap}) \Bigg) }{R_{tpap} + R_{rv}} &\qquad P_{rv} > P_{pap1} \\
    P_{pap2} = \dfrac{- R_{rv} F_{pap} R_{tpap} + \Bigg( R_{rv} \dfrac{V_{pap} - V_{pap0}}{C_{pap}} - \psi (V_{pap}) \Bigg) }{R_{rv}} &\qquad P_{rv} \leq P_{pap1}
    \end{dcases}
\end{equation}

\begin{equation}
    P_{pad} = F_{pap} R_{tpad} - F_{pad} R_{tpad} + \frac{V_{pad} - V_{pad0}}{C_{pad}} - \psi (V_{pad})
\end{equation}

\begin{equation} 
    V_{p,i} = \frac{V_{p,i} - V_{p,i,0}}{C_{p,i}} - \psi (V_{p,i})
\end{equation}

\begin{equation}
    F_{ps} = \frac{P_{pa} - P_{pv}}{R_{ps}}
\end{equation}

\begin{equation}
    F_{pa} = \frac{P_{pa} - P_{pc}}{R_{pa}}
\end{equation}

\begin{equation}
    F_{pc} = \frac{P_{pc} - P_{pv}}{R_{pc}}
\end{equation}

\begin{equation}
    F_{pv} = \frac{P_{pv} - P_{pa}}{R_{pv}}
\end{equation}

\begin{equation}
    \frac{dF_{pap}}{dt} = \frac{ P_{pap} - P_{pad} - F_{pap} R_{pap} }{L_{pap}}
\end{equation}

\begin{equation}
    \frac{dF_{pad}}{dt} = \frac{ P_{pad} - P_{pa} - F_{pad} R_{pad} }{L_{pad}}
\end{equation}

\begin{equation}
    \frac{dV_{pad}}{dt} = F_{pap} - F_{pad}
\end{equation}

\begin{equation}
    \frac{dV_{pap}}{dt} = F_{rv} - F_{pap}
\end{equation}

\begin{equation}
    \frac{dV_{pa}}{dt} = F_{pad} - F_{ps} - F_{pa}
\end{equation}

\begin{equation}
    \frac{dV_{pc}}{dt} = F_{pa} - F_{pc}
\end{equation}

\begin{equation} \label{eq:dvpvdt}
    \frac{dV_{pv}}{dt} = F_{pc} + F_{ps} - F_{pv}
\end{equation}

\subsubsection{Coronary circulation}

\begin{equation} \label{eq:pcorepi}
    P_{corepi} = P_{aop}
\end{equation}

\begin{equation}
    P_{corintra} = \frac{V_{corintra} - V_{corintra0}}{C_{corintra}} - \psi (V_{})
\end{equation}

\begin{equation}
    P_{corcap} = \frac{V_{corcap} - V_{corcap0}}{C_{corcap}} - \psi (V_{corcap})
\end{equation}

\begin{equation}
    P_{corvn} = \frac{V_{corvn} - V_{corvn0}}{C_{corvn}} - \psi (V_{corvn})
\end{equation}

\begin{equation}
    P_{im} = \Bigg\lfloor \frac{P_{lv}}{2} \Bigg\rfloor
\end{equation}

\begin{equation}
    P_{corintrac} = P_{corintra} + P_{im}
\end{equation}

\begin{equation}
    P_{corcapc} = P_{corcap} + P_{im}
\end{equation}

\begin{equation}
    P_{corvnc} = P_{corvn}
\end{equation}

\begin{equation}
    F_{corepi} = \frac{P_{corepi} - P_{corintrac}}{R_{corepi}}
\end{equation}

\begin{equation}
    F_{corintra} = \frac{P_{corintrac} - P_{corcapc}}{R_{corintra}}
\end{equation}

\begin{equation}
    F_{corcap} = \frac{P_{corcapc} - P_{corvnc}}{R_{corcap}}
\end{equation}

\begin{equation}
    F_{corvn} = \frac{P_{corvnc} - P_{ra}}{R_{corvn}}
\end{equation}

\begin{equation}
    \frac{dV_{corepi}}{dt} = F_{lv} - \frac{dV_{vaop}}{dt} - F_{aop} - F_{corepi}
\end{equation}

\begin{equation}
    \frac{dV_{corintra}}{dt} = F_{corepi} - F_{corintra}
\end{equation}

\begin{equation}
    \frac{dV_{corcap}}{dt} = F_{corintra} - F_{corcap}
\end{equation}

\begin{equation} \label{eq:dvcorvndt}
    \frac{dV_{corvn}}{dt} = F_{corcap} - F_{corvn}
\end{equation}

\subsubsection{Baroreceptor}

\begin{equation} \label{eq:bvaso}
    b_{vaso} = 1 - a_{vaso}
\end{equation}

\begin{equation}
    f_{vaso} = a_{vaso} + \dfrac{b_{vaso}}{e^{\tau_{vaso} (N_{vaso}-N_{o,vaso}})) + 1}
\end{equation}

\begin{equation} \label{eq:dnbrdt}
    N_{br,t} = \frac{dN_{br}}{dt}
\end{equation}

\begin{equation} \label{eq:d2nbrdt2}
    \frac{dN_{br,t}}{dt} = \frac{-(a_2 + a) N_{br,t} - N_{br} + K\Bigg( ABP_{shift} + a_1 \dfrac{dABP_{fol}}{dt} \Bigg) }{a_2 a}
\end{equation}

\begin{equation} \label{eq:dnidt} 
    \frac{dN_i}{dt} = 
    \begin{dcases}
    \dfrac{-N_i + K_i N_{br} (t - l_i)}{T_i} &\qquad t - t_{HB} (0) > l_i \\
    0 &\qquad t - t_{HB} (0) \leq l_i
    \end{dcases}
\end{equation}

\subsubsection{Blood volumes}

\begin{equation} \label{eq:vcorcic}
    V_{corcic} = V_{corepi} + V_{corintra} + V_{corcap} + V_{corvn}
\end{equation}

\begin{equation}
    V_{heart} = V_{ra} + V_{rv} + V_{la} + V_{lv} + V_{corcic}
\end{equation}

\begin{equation}
    V_{sysart} = V_{aop} + V_{aod} + V_{sap} + V_{sa}
\end{equation}

\begin{equation}
    V_{sysven} = V_{sv} + V_{vc}
\end{equation}

\begin{equation}
    V_{pulart} = V_{pap} + V_{pad} + V_{pa}
\end{equation}

\begin{equation} \label{eq:tbv}
    TBV = V_{heart} + V_{sysart} + V_{sc} + V_{sysven} + V_{pulart} + V_{pc} + V_{pv}
\end{equation}

\begin{landscape}

\subsection{Parameters of the renin-angiotensin system}

\begin{table}[!ht]
\renewcommand{\arraystretch}{1.5}
\centering
\caption{Pharmacokinetic parameters}
\label{tab:pk}
\begin{tabular}{llllll}
\hline
\textbf{Parameter} & \textbf{NRF} & \textbf{IRF} & \textbf{Units} & \textbf{Sources} & \textbf{Description} \\ \hline
$k_a$ & 1.907 & 1.645 & $h^{-1}$ & \cite{versypt2017pharmacokinetic,shionoiri1992pharmacokinetics} & absorption rate constant \\
$k_e$ & $1.33 \times 10^{-1}$ & $3.45 \times 10^{-2}$ & $h^{-1}$ & \cite{versypt2017pharmacokinetic,shionoiri1992pharmacokinetics} & elimination rate constant \\
$V/F$ & $7.09 \times 10^4$ & $1.07 \times 10^5$ & $mL$ & \cite{versypt2017pharmacokinetic,shionoiri1992pharmacokinetics} & ratio of the volume of distribution to the fraction of the drug absorbed \\ 
\hline
\end{tabular}
\end{table}

\renewcommand{\arraystretch}{1.5}
\begin{longtable}[c]{lllll}
\caption{Pharmacodynamic parameters}
\label{tab:pd}\\

\hline
\textbf{Parameter} & \textbf{Value} & \textbf{Units} & \textbf{Sources} & \textbf{Description}  \\
\hline
\endfirsthead
\multicolumn{5}{c}%
{{\bfseries Table \thetable\ continued from previous page}} \\
\hline
\textbf{Parameter} & \textbf{Value} & \textbf{Units} & \textbf{Sources} & \textbf{Description}  \\
\hline \\
\endhead

$k_{AGT}$ & $2.27 \times 10^6$ & $nmol/L/h$ & \cite{pilvankar2018mathematical,pilvankar2019glucose} & constant production rate of AGT \\
$k_{Renin}$ & $6.44 \times 10^4$ & $h^{-1}$ & \cite{versypt2017pharmacokinetic,pilvankar2019glucose} & ANG-I production rate due to renin \\
$k_{NEP}$ & $0.583$ & $h^{-1}$ & \cite{pilvankar2018mathematical,pilvankar2019glucose} & NEP-catalyzed conversion rate from ANG-I to ANG-(1-7) \\
$k_{AT2}$ & $25.1$ & $h^{-1}$ & \cite{pilvankar2018mathematical,pilvankar2019glucose} & rate parameter for binding of ANG II to AT1R \\
$k_{APA}$ & $43.6$ & $h^{-1}$ & \cite{pilvankar2018mathematical,pilvankar2019glucose} & APA-catalyzed conversion rate from ANG-II to ANG-III  \\
$h_{AGT}$ & $10.0$ & $h$ & \cite{lo2011using,pilvankar2018mathematical,pilvankar2019glucose} & AGT half-life degradation rate \\
$h_{Renin}$ & $0.250$ & $h$ & \cite{pilvankar2018mathematical,pilvankar2019glucose} & renin half-life degradation rate \\
$h_{ANG I}$ & $1.72 \times 10^{-4}$ & $h$ & \cite{lo2011using,pilvankar2018mathematical,pilvankar2019glucose} & ANG-I half-life degradation rate \\
$h_{ANG II}$ & $5 \times 10^{-3}$ & $h$ & \cite{lo2011using,pilvankar2018mathematical,pilvankar2019glucose} & ANG-II half-life degradation rate \\
$[Drug]_{50}$ & $2.20$ & $ng/mL$ & \cite{shionoiri1992pharmacokinetics,pilvankar2018mathematical,pilvankar2019glucose} & drug concentration yielding 50\% inhibition \\
$m$ & $0.99$ & - & \cite{shionoiri1992pharmacokinetics,pilvankar2018mathematical,pilvankar2019glucose} & degree of sigmoidicity of the Hill function \\
$a_{Renin}$ & $5.47 \times 10^{-4}$ & $L/mmol/h$ & \cite{pilvankar2018mathematical,pilvankar2019glucose} & slope of the linear dependence of renin from glucose \\
$b_{Renin}$ & $6.16 \times 10^{-11}$ & $h^{-1}$ & \cite{pilvankar2018mathematical,pilvankar2019glucose} & intercept of the linear dependence of renin from glucose \\
$a_{ACE}$ & $0.889$ & $L/mmol/h$ & \cite{pilvankar2018mathematical,pilvankar2019glucose} & slope of the linear dependence of ACE from glucose \\
$b_{ACE}$ & $163$ & $h^{-1}$ & \cite{pilvankar2018mathematical,pilvankar2019glucose} & intercept of the linear dependence of ACE from glucose \\
$a_{AT1}$ & $2.55$ & $L/mmol/h$ & \cite{pilvankar2018mathematical,pilvankar2019glucose} & slope of the linear dependence of AT1R from glucose \\
$b_{AT1}$ & $464$ & $h^{-1}$ & \cite{pilvankar2018mathematical,pilvankar2019glucose} & intercept of the linear dependence of AT1R from glucose \\
$k_{f,sys}$ & $6.25 \times 10^{-2}$ & $h^{-1}$ & \cite{versypt2017pharmacokinetic,pilvankar2019glucose} & ANG-II feedback parameter on renin \\
$f_{sys}$ & $0.397$ & $nmol/L$ & \cite{versypt2017pharmacokinetic,pilvankar2019glucose} & ANG-II feedback parameter on renin \\
$[AGT]_0$ & $1.70 \times 10^7$ & $nmol/L$ & \cite{pilvankar2018mathematical,pilvankar2019glucose} & AGT initial concentration \\
$[Renin]_0$ & $2.06 \times 10^{-4}$ & $nmol/L$ & \cite{versypt2017pharmacokinetic,pilvankar2019glucose} & renin initial concentration \\
$[ANG I]_0$ & $271$ & $nmol/L$ & \cite{pilvankar2018mathematical,pilvankar2019glucose} & ANG-I initial concentration \\
$[ANG II]_0$ & $21.0$ & $nmol/L$ & \cite{pilvankar2018mathematical,pilvankar2019glucose} & ANG-II initial concentration \\
$[ANG II]_{0,sys \text{NRF}}$ & $1.65 \times 10^{-2}$ & $nmol/L$ & \cite{versypt2017pharmacokinetic,pilvankar2019glucose} & systemic ANG-II initial concentration for normal renal individuals \\
$[ANG II]_{0,sys \text{IRF}}$ & $2.05 \times 10^{-2}$ & $nmol/L$ & \cite{versypt2017pharmacokinetic,pilvankar2019glucose} & systemic ANG-II initial concentration for impaired renal individuals \\
$h_{ANG17}$ & $0.5$ & $h$ & \cite{lo2011using} & ANG-17 half-life degradation rate \\
$[ANG17]_0$ & $9.858$ & $nmol/L$ & \cite{hisatake2017serum} & ANG-17 initial concentration \\
$h_{AT1R}$ & $0.2$ & $h$ & \cite{lo2011using} & AT1R half-life degradation rate \\
$[AT1R]_0$ & $16.2$ & $nmol/L$ & \cite{hisatake2017serum} & AT1R initial concentration \\
$h_{AT2R}$ & $0.2$ & $h$ & \cite{lo2011using} & AT2R half-life degradation rate \\
$[AT2R]_0$ & $5.4$ & $nmol/L$ & \cite{hisatake2017serum} & AT2R initial concentration \\
$k_{ACE2,0}$ & $0.385$ & $h^{-1}$ & \cite{pilvankar2018mathematical,pilvankar2019glucose} & ACE-catalyzed conversion rate from ANG-II to ANG-(1-7) \\
$s_I$ & $0.1$ & $h^{-1} L/nmol$ & - & severity of SARS-CoV-2 infection \\
$e_{AI}$ & $0.347$ & - & - & efficiency of anti-inflammatory pathways \\
\hline
\end{longtable}

\subsection{Parameters of the diabetic model}

\renewcommand{\arraystretch}{1.5}
\begin{longtable}[c]{lllll}
\caption{Diabetic parameters}
\label{tab:diabetes}\\

\hline
\textbf{Parameter} & \textbf{Value} & \textbf{Units} & \textbf{Sources} & \textbf{Description}  \\
\hline
\endfirsthead
\multicolumn{5}{c}%
{{\bfseries Table \thetable\ continued from previous page}} \\
\hline
\textbf{Parameter} & \textbf{Value} & \textbf{Units} & \textbf{Sources} & \textbf{Description}  \\
\hline \\
\endhead

$k$ & $432$ & $d^{-1}$ & \cite{toffolo1980quantitative} & combined insulin uptake at the liver, kidneys, and insulin receptors \\
$\alpha$ & $20000$ & $mg^2 dl^{-2}$ & \cite{malaisse1967new} & glucose concentration yielding 50\% of insulin secretion \\
$\sigma$ & $43.2$ & $\mu U \ ml^{-1}d^{-1}$ & \cite{bergman1981physiologic,malaisse1967new,toffolo1980quantitative} & maximal rate secretion of insulin by $\beta$ cells \\
$R_0$ & $864$ & $mg \ dl^{-1}d^{-1}$ & \cite{bergman1981physiologic,finegood1997application} & net rate of production at zero glucose \\
$R_1$ & $1$ & $ml\ dl^{-1}\ g^{-1}$ & - & net rate of glucose increase due to meals \\
$R_2$ & $0.1$ & $ml\ dl^{-1}\ kcal^{-1}$ & - & net rate of glucose consumption due to workouts \\
$EG_0$ & $0.44$ & $d^{-1}$ & \cite{bergman1981physiologic,finegood1997application} & total glucose effectiveness at zero insulin \\
$SI$ (normal) & $1.62$ & $ml \ \mu U^{-1}d^{-1}$ & \cite{finegood1997application} & normal insulin sensitivity \\
$SI$ (diabetic) & $0.52$ & $ml \ \mu U^{-1}d^{-1}$ & \cite{finegood1997application} & diabetic insulin sensitivity \\
$r_0$ & $0.06$ & $d^{-1}$ & \cite{bergman1981physiologic,imamura1988severe,finegood1997application} & death rate at zero glucose \\
$r_1$ & $0.00084$ & $mg^{-1} dl\ d^{-1}$ & \cite{bergman1981physiologic,imamura1988severe,finegood1997application} & I-order coefficient for $\beta$ cell replication \\
$r_2$ & $0.0000024$ & $mg^{-2} dl\ d^{-1}$ & \cite{bergman1981physiologic,imamura1988severe,finegood1997application} & II-order coefficient for $\beta$ cell replication \\
$i_0$ & $87$ & - & \cite{Topp2000} & insulin resistance self-inhibition rate \\
$m$ & $2$ & - &  \cite{Topp2000} & insulin resistance progression rate due to pro-inflammatory cytokines \\
$q$ & $0.017$ & $ml / \mu U$ &  \cite{Topp2000} & insulin resistance progression rate due to insulin concentration \\
$I_0$ & $13.59$ & $\mu U / ml$ &  \cite{Topp2000} & initial insulin concentration \\
$G_0$ & $100$ & $ml/dl$ &  \cite{Topp2000} & initial glucose concentration \\
$\beta_{f,0}$ & $407.73$ & - &  \cite{Topp2000} & number of functional $\beta$-cells \\
$I_{R,0}$ & $0.359$ & - &  - & initial insulin resistance \\
\hline
\end{longtable}

\subsection{Parameters of the stiffness model}

\renewcommand{\arraystretch}{1.5}
\begin{longtable}[c]{lllll}
\caption{Stiffness parameters}
\label{tab:stiff}\\

\hline
\textbf{Parameter} & \textbf{Value} & \textbf{Units} & \textbf{Sources} & \textbf{Description}  \\
\hline
\endfirsthead
\multicolumn{4}{c}%
{{\bfseries Table \thetable\ continued from previous page}} \\
\hline
\textbf{Parameter} & \textbf{Value} & \textbf{Units} &  & \textbf{Description}  \\
\hline \\
\endhead

$k_{SARS}$ & $0.15$ & $h$ & inflammation rate due to SARS-CoV-2 \\
$k_D$ & $0.001$ & $mL/ng$ & inflammation rate due to ACEi surplus \\
$k_G$ & $0.1$ & $L/mmol$ & inflammation rate due to glucose surplus \\
$k_{eff}$ (healthy state) & $0.035$ & - & anti-inflammatory response rate \\
$k_{eff}$ (during infection) & $0.693$ & - & anti-inflammatory response rate \\
$\beta_1$ & $0.006$ & - & compliance reduction rate due to ageing \\
$\beta_0$ & $1.2$ & - & compliance reduction intercept due to ageing \\
$\text{IR}_0$ & $0.385$ & - & inflammatory response initial condition \\
$\beta_h$ & $0.0008$ & $mmHg\ mL\ U^{-1}$ & heparin impact on blood pressure \\
$\beta_D$ & $0.05$ & $mmHg\ mL\ ng^{-1}$ & vitamin D impact on blood pressure \\
$[heparin]_0$ & $5000$ & $U\ ml^{-1}$ & initial heparin dose for intra venous continuous infusion treatment \\
$[D]_0$ & $30$ & $ng\ mL^{-1}$ & vitamin D recommended concentration \\
\hline
\end{longtable}

\subsection{Parameters of the open-loop circulatory model}

\renewcommand{\arraystretch}{1.5}
\begin{longtable}[c]{llll}
\label{tab:cardio}\\

\hline
\textbf{Parameter} & \textbf{Value} & \textbf{Units} & \textbf{Description} \\
\hline
\endfirsthead
\multicolumn{4}{c}%
{{\bfseries Table \thetable\ continued from previous page}} \\
\hline
\textbf{Parameter} & \textbf{Value} & \textbf{Units} & \textbf{Description}\\
\hline
\endhead
$Ts1v$ & $0.349$ & $sec$ & Scaler to set ventricular systolic fraction of heart cycle \\
$Ts1a$ & $0.2$ & $sec$ & Scaler to set atrial systolic fraction of heart cycle \\
$Ts2$ & $1$ & $hz$ & Unit balance scalar for Tsa and Tsv functions \\
$offv$ & $0.0263$ & $sec$ & Parameter to match model and measured end-diastolic ABP \\
$Vlvd0$ & $71.816$ & $ml$ & Unstressed end-diastolic left ventricle volume \\
$Vlvs0$ & $23.699$ & $ml$ & Unstressed end-systolic left ventricle volume \\
$Vrvd0$ & $102.881$ & $ml$ & Unstressed end-diastolic right ventricle volume \\
$Vrvs0$ & $53.498$ & $ml$ & Unstressed end-systolic right ventricle volume \\
$Vlad0$ & $70$ & $ml$ & Unstressed end-diastolic left atrium volume \\
$Vlas0$ & $40$ & $ml$ & Unstressed end-systolic left atrium volume \\
$Vrad0$ & $60$ & $ml$ & Unstressed end-diastolic right atrium volume \\
$Vras0$ & $53$ & $ml$ & Unstressed end-systolic right atrium volume \\
$Rra$ & $0.001$ & $mmHg \ s \ ml^{-1}$ & Tricuspid valve resistance \\
$Rla$ & $0.001$ & $mmHg \ s \ ml^{-2}$ & Mitral valve resistance \\
$Rlv$ & $0.0001$ & $mmHg \ s \ ml^{-3}$ & Aortic valve resistance \\
$Rrv$ & $0.0001$ & $mmHg \ s \ ml^{-4}$ & Pulmonary valve resistance \\
$PRint$ & $0.12$ & $sec$ & Difference in atrial, venticular activation times \\
$KElv$ & $1$ &  & Scaling factor for maximum left ventricular elastance \\
$KErv$ & $1$ &  & Scaling factor for maximum right ventricular elastance \\
$Emaxlv1$ & $5.4$ & $mmHg/ml$ & Maximum elastance of first left ventricle component \\
$Eminlv$ & $0.09$ & $mmHg/ml$ & Minimum elastance of first left ventricle component \\
$Emaxrv1$ & $0.53$ & $mmHg/ml$ & Maximum elastance of first right ventricle component \\
$Eminrv$ & $0.0343$ & $mmHg/ml$ & Minimum elastance of first right ventricle component \\
$EDVLV$ & $125.993$ & $ml$ &  \\
$EDVRV$ & $175.865$ & $ml$ &  \\
$Emaxra$ & $0.13$ & $mmHg/ml$ & Maximum elastance right ventricle \\
$Eminra$ & $0.085$ & $mmHg/ml$ & Minimum elastance left ventricle \\
$Emaxla$ & $0.299$ & $mmHg/ml$ & Maximum elastance right ventricle \\
$Eminla$ & $0.185$ & $mmHg/ml$ & Minimum elastance left ventricle \\
$KCOMAP$ & $3$ & $L/mmHg/min^2$ &  \\
$Raop$ & $0.0001$ & $mmHg \ sec \ ml^{-1}$ & Proximal aortic resistance \\
$Rtaop$ & $0.02$ & $mmHg \ sec \ ml^{-1}$ & Transmural proximal aortic resistance \\
$Rcrb$ & $6.8284$ & $mmHg \ sec \ ml^{-1}$ & Cerebral circulation resistance \\
$Raod$ & $0.0129$ & $mmHg \ sec \ ml^{-1}$ & Distal aortic resistance \\
$Rtaod$ & $1$ & $mmHg \ sec \ ml^{-1}$ & Transmural distal aortic resistance \\
$Rsap$ & $0.003$ & $mmHg \ sec \ ml^{-1}$ & Systemic arteriolar resistance \\
$Rsc$ & $0.155$ & $mmHg \ sec \ ml^{-1}$ & Systemic capillaries resistance \\
$Rsv$ & $0.138$ & $mmHg \ sec \ ml^{-1}$ & Systemic veins resistance \\
$Caop$ & $0.263$ & $ml \ mmHg^{-1}$ & Aortic proximal compliance \\
$Caod$ & $0.639$ & $ml \ mmHg^{-1}$ & Aortic distal compliance \\
$Csap$ & $1.482$ & $ml \ mmHg^{-1}$ & Systemic arterioles compliance \\
$Csc$ & $5.767$ & $ml \ mmHg^{-1}$ & Systemic capillaries compliance \\
$Vaop0$ & $9.520$ & $ml$ & Proximal aorta unstressed volume \\
$Vaod0$ & $23.11$ & $ml$ & Distal aorta unstressed volume \\
$Vsap0$ & $52.94$ & $ml$ & Systemic arteries unstressed volume \\
$Vsc0$ & $71.02$ & $ml$ & Systemic capillaries unstressed volume \\
$Laop$ & $1e-05$ & $mmHg \ sec^2 \ ml^{-1}$ & Proximal aorta intertance \\
$Laod$ & $2e-05$ & $mmHg \ sec^2 \ ml^{-1}$ & Distal aorta inertance \\
$Kc$ & $497.785$ & $mmHg$ & Active vasomotor tone scaling parameter for systemic arterial pressure \\
$Do$ & $50$ & $ml$ & Active vasomotor tone volume parameter for systemic arterial pressure \\
$Vsa0$ & $485.762$ & $ml$ & Minimal volume of systemic arteries \\
$Vsa\_max$ & $577.711$ & $ml$ & Maximal luminal volume of systemic arteries \\
$Kp1$ & $0.0299$ & $mmHg$ & Passive vasomotor tone scaling parameter for systemic arterial pressure \\
$Kp2$ & $0.05$ & $mmHg \ ml^{-2}$ & Passive vasomotor tone scaling parameter for systemic arterial pressure \\
$Kr$ & $0.01$ & $mmHg \ sec \ ml^{-1}$ & Pressure scaling constant for systemic arterial resistance \\
$Rsa0$ & $0.581$ & $mmHg \ sec/ml$ & Offset parameter for systemic arteriolar resistance \\
$tau\_p$ & $0.1$ & $ml^{-1}$ & Passive vasomotor tone constant for systemic arterial pressure \\
$Ksv$ & $0.74$ &  & Scaling factor used to optimize systemic venous pressure-volume relationship \\
$Kv1$ & $30.21$ & $mmHg$ & Scaling factor for systemic venous pressure \\
$Vmax\_sv$ & $3379.55$ & $ml$ & Maximal volume of lumped systemic veins \\
$D2$ & $-5$ & $mmHg$ & Offsetting constant for partially collapsed Vena cava pressure \\
$K1$ & $0.046$ & $mmHg \ ml^{-1}$ & Scaling factor for Vena cava PV relationship \\
$K2$ & $0.374$ & $mmHg$ & Scaling factor for partially collapsed Vena cava pressure \\
$KR$ & $0.001$ & $mmHg \ sec \ ml^{-1}$ & Scaling factor for Vena cava resistance \\
$R0$ & $0.025$ & $mmHg \ sec \ ml^{-1}$ & Vena cava resistance offset parameter \\
$Vvc0$ & $129.649$ & $ml$ & Unstressed volume of Vena cava \\
$Vmax\_vc$ & $350.53$ & $ml$ & Maximum volume of Vena cava \\
$Vmin\_vc$ & $50.01$ & $ml$ & Minimum volume of Vena cava \\
$tauCO$ & $15$ & $sec$ & Cardiac output equation time constant \\
$Kxp$ & $2$ & $mmHg$ & P-V curve shaping parameter \\
$Kxv$ & $8$ & $ml$ & P-V curve shaping parameter \\
$Kxv1$ & $1$ & $ml$ & P-V curve shaping parameter \\
$Kxp1$ & $1$ & $mmHg$ & P-V curve shaping parameter \\
$tauMAP$ & $2$ & $sec$ & Time constant for mean arterial pressure ODE \\
$tauABP$ & $0.001$ & $sec$ & Time constant for ABP follower \\
$Rtpap$ & $0.1$ & $mmHg \ sec \ ml^{-1}$ & Proximal pulmonary arterial transmural resistance \\
$Rtpad$ & $0.2$ & $mmHg \ sec \ ml^{-1}$ & Distal pulmonary arterial transmural resistance \\
$Rpap$ & $0.0001$ & $mmHg \ sec \ ml^{-1}$ & Proximal pulmonary resistance \\
$Rpad$ & $0.0299$ & $mmHg \ sec \ ml^{-1}$ & Distal proximal pulmonary resistance \\
$Rps$ & $4.333$ & $mmHg \ sec \ ml^{-1}$ & Pulmonary shunt resistance \\
$Rpa$ & $0.057$ & $mmHg \ sec \ ml^{-1}$ & Pulmonary arterioles resistance \\
$Rpc$ & $0.032$ & $mmHg \ sec \ ml^{-1}$ & Pulmonary capillaries resistance \\
$Rpv$ & $0.0001$ & $mmHg \ sec \ ml^{-1}$ & Pulmonary veins resistance \\
$Cpap$ & $1.445$ & $ml \ mmHg^{-1}$ & Proximal pulmonary arterial compliance \\
$Cpad$ & $2.531$ & $ml \ mmHg^{-1}$ & Distal pulmonary arterial compliance \\
$Cpa$ & $3.102$ & $ml \ mmHg^{-1}$ & Pulmonary arterioles compliance \\
$Cpc$ & $9.117$ & $ml \ mmHg^{-1}$ & Pulmonary capillaries compliance \\
$Cpv$ & $52.267$ & $ml \ mmHg^{-1}$ & Pulmonary veins compliance \\
$Vpap0$ & $9.81$ & $ml$ & Proximal pulmonary artery unstressed volume \\
$Vpad0$ & $17.16$ & $ml$ & Distal pulmonary artery unstressed volume \\
$Vpa0$ & $17.16$ & $ml$ & Small pulmonary arteries unstressed volume \\
$Vpc0$ & $29.42$ & $ml$ & Pulmonary capillaries unstressed volume \\
$Vpv0$ & $29.597$ & $ml$ & Pulmonary veins unstressed volume \\
$Lpap$ & $0.00018$ & $mmHg \ sec^2 \ ml^{-1}$ & Proximal arterial inertance \\
$Lpad$ & $0.00019$ & $mmHg \ sec^2 \ ml^{-1}$ & Distal pulmonary artery inertance \\
$Rcorepi$ & $5.285$ & $s \ ml^-1 \ mmHg$ & Proximal epicardial arteries resistance \\
$Rcorintra$ & $10.147$ & $s \ ml^-1 \ mmHg$ & Distal epicardial arteries resistance \\
$Rcorcap$ & $4.228$ & $s \ ml^-1 \ mmHg$ & Coronary capillaries resistance \\
$Rcorvn$ & $1.48$ & $s \ ml^-1 \ mmHg$ & Small coronary veins resistance \\
$Ccorepi$ & $0.074$ & $ml/mmHg$ & Compliance of proximal epicardial arteries \\
$Ccorintra$ & $0.134$ & $ml/mmHg$ & Compliance of distal epicardial arteries \\
$Ccorcap$ & $0.94$ & $ml/mmHg$ & Compliance of coronary capillaries \\
$Ccorvn$ & $2.45$ & $ml/mmHg$ & Compliance of small coronary veins \\
$Vcorepi0$ & $2.69$ & $ml$ & Epicardial arteries unstressed volume \\
$Vcorintra0$ & $2.685$ & $ml$ & Intramyocardial arteries unstressed volume \\
$Vcorcap0$ & $2.523$ & $ml$ & Coronary capillaries unstressed volume \\
$Vcorvn0$ & $2.493$ & $ml$ & Coronary veins unstressed volume \\
$a$ & $0.001$ & $sec$ & Time constant for baroreceptor firing rate \\
$a1$ & $0.036$ & $sec$ & Time constant for baroreceptor firing rate \\
$a2$ & $0.0018$ & $sec$ & Time constant for baroreceptor firing rate \\
$K$ & $0.991$ & $sec^{-1} \ mmHg^{-1}$ & Baroreceptor gain (used to account for units) \\
$K\_con$ & $1$ &  & CNS gain for contractility control \\
$T\_con$ & $10$ & $sec$ & CNS time parameter for contractility control \\
$l\_con$ & $3$ & $sec$ & CNS time delay for contractility control \\
$a\_con$ & $0.299$ &  & Time constant for efferent contractility firing \\
$b\_con$ & $0.699$ &  & Time constant for efferent sympathetic contractility firing \\
$tau\_con$ & $0.04$ & $sec$ & Time parameter for efferent sympathetic contractility firing \\
$No\_con$ & $110$ & $sec^{-1}$ & Frequency parameter for efferent sympathetic contractility firing \\
$K\_vaso$ & $1$ &  & CNS gain for vasomotor tone control \\
$T\_vaso$ & $6$ & $sec$ & CNS time parameter for vasomotor tone control \\
$l\_vaso$ & $3$ & $sec$ & CNS time delay for vasomotor tone control \\
$a\_vaso$ & $-0.466$ &  & Time constant for efferent vasomotor tone firing \\
$tau\_vaso$ & $0.04$ & sec & Time parameter for efferent vasomotor tone firing \\
$No\_vaso$ & $110$ & $sec^{-1}$ & Frequency parameter for efferent vasomotor tone firing \\
$amin$ & $-2.806$ &  & Contractility control offset \\
$bmin$ & $0.699$ &  & Contractility control offset \\
$Ka$ & $5$ &  & Contractility control scaling factor \\
$Kb$ & $0.5$ &  & Contractility control scaling factor \\
\hline
\end{longtable}

\end{landscape}

\end{document}